\documentclass[11pt,psfig]{article}
\pdfoutput=1
\usepackage{graphicx}
\usepackage{subfigure,amssymb,amsmath,slashed}
\usepackage{cite,color,xcolor,url,cancel}
\usepackage{bm}
\usepackage{float}
\usepackage{multirow}
\usepackage{tikz}
\usepackage{multicol}
\usepackage{mathtools}
\usepackage{tikz-feynman}
\usetikzlibrary{shapes.geometric, arrows}
\usetikzlibrary{calc,patterns,angles,quotes}
\usepackage[customcolors,norndcorners]{hf-tikz}
\usepackage{jheppub}
\usepackage{multirow, makecell}

\newcommand{\be}{\begin{equation}}
\newcommand{\ee}{\end{equation}}
\newcommand{\bea}{\begin{eqnarray}}
\newcommand{\eea}{\end{eqnarray}}
\newcommand{\noi}{\noindent}
\newcommand{\bse}{\begin{subequations}}
\newcommand{\ese}{\end{subequations}}
\newcommand{\tp}{T^\prime}
\newcommand{\nn}{\nonumber}
\newcommand{\ra}{\rightarrow}
\newcommand{\zp}{Z^\prime}
\newcommand{\rp}{\rho^\prime}
\newcommand{\xp}{x^\prime}
\newcommand{\zmt}{Z_{\mu\tau}}
\newcommand{\x}{\chi}
\newcommand{\mt}{\mu \tau}
\newcommand{\lmt}{L_\mu - L_\tau}
\newcommand{\smgauge}{SU(3)_c \otimes SU(2)_L \otimes U(1)_Y}
\newcommand{\eps}{\epsilon}

\title{Non-adiabatic evolution of dark sector in the 
presence of $U(1)_{L_\mu - L_\tau}$ gauge symmetry}

\author[a]{Ananya Tapadar,}
\affiliation[a]{School of Physical Sciences, Indian Association 
for the Cultivation of Science,\\  
	2A \& 2B Raja S.C. Mullick Road, Jadavpur, 
	Kolkata 700 032, India}

\author[a]{Sougata Ganguly,}
\author[a]{Sourov Roy}

\emailAdd{intat@iacs.res.in}
\emailAdd{tpsg4@iacs.res.in}
\emailAdd{tpsr@iacs.res.in}

\abstract{
In secluded dark sector scenario, the connection between
the visible and the dark sector can be established through 
a portal coupling and its presence opens up the possibility
of non-adiabatic evolution of the dark sector.
To study the non-adiabatic evolution of the dark sector, 
we have considered a $U(1)_{L_\mu - L_\tau} \otimes U(1)_X$
extension of the standard model (SM). Here
the dark sector is charged only under $U(1)_X$ gauge symmetry whereas 
the SM fields are singlet under this symmetry. Due to the
presence of tree-level
kinetic mixing between $U(1)_X$ and $U(1)_{L_\mu - L_\tau}$
gauge bosons, the dark sector evolves non-adiabatically
and thermal equilibrium between the visible and dark sector
is governed by the portal coupling.
Depending on the values of the portal coupling ($\epsilon$), 
dark sector gauge coupling ($g_X$),
mass of the dark matter ($m_\chi$) and 
mass of the dark vector boson ($m_{Z^\prime}$), we study the temperature 
evolution of the dark sector as well as the various non-equilibrium 
stages of the dark sector in detail.
Furthermore we have also investigated the constraints
on the model parameters from various laboratory and astrophysical searches. We 
have found that the parameter space for the non-adiabatic
evolution of dark sector is significantly constrained for 
$m_{Z^\prime}$ $\lesssim 100 \, {\rm MeV}$ from
the observations of beam dump experiments, stellar cooling etc. 
The relic density satisfied region of our parameter space
is consistent with the bounds from direct detection, and
self interaction of dark matter (SIDM) for the mass ratio
$r \equiv  m_{Z^\prime}/m_\x = 10^{-3}$ and these bounds will be more relaxed 
for larger values of $r$. However the constraints from
measurement of diffuse $\gamma$-ray background flux
and cosmic microwave background (CMB) anisotropy
are strongest for $r = 10^{-1}$ and for smaller 
values of $r$, they are not significant.
}
\vspace{1cm}
\graphicspath{{figures/}}
\begin{document}
\maketitle
\section{Introduction}
Overwhelming astrophysical and cosmological evidences establish
the fact that about one fourth of our Universe is made of some unknown
non-luminous matter which is known as Dark Matter (DM). 
The precise value of the abundance of the DM is measured by the
satellite borne experiments such as WMAP \cite{Hinshaw:2012aka} and 
Planck \cite{Aghanim:2018eyx} and it is given by 
$\Omega_{\rm DM} h^2 = 0.1200 \pm 0.0012 $ \cite{Aghanim:2018eyx}. 
There are other indirect evidences of the 
presence of DM such as Bullet cluster observations \cite{Clowe:2006eq},
rotation curve of the spiral galaxies \cite{Sofue:2000jx}, 
gravitational lensing of the distant objects \cite{Bartelmann:1999yn} etc.
In the face of several (in)direct evidences,
the origin and nature of DM is elusive to date and 
our knowledge on DM physics is mostly confined to its gravitational
interaction.

Over last two decades, people have extensively studied weakly
interacting massive particle (WIMP) as a well motivated DM candidate
\cite {Gondolo:1990dk,Srednicki:1988ce,Bertone:2004pz,Feng:2010gw}.
In this paradigm it is assumed that the DM is in thermal equilibrium
with the Standard Model (SM) particles
in the early Universe and its coupling strength with
the SM particles is of the same order as electroweak (EW) coupling.
It was found that for correct relic
abundance the mass of the DM should be in the EW scale. This remarkable
result is known as ``WIMP miracle" \cite{Feng:2010gw}. 
Despite the fact that its coupling with the SM fields lies
in the EW scale, there are no
positive signals in detecting DM in astrophysical
and laboratory experiments \cite{Lin:2019uvt,Roszkowski:2017nbc,Arcadi:2017kky}.

Motivated by the null results of these experiments, people have suggested
many other possibilities for the  DM candidate such as 
feebly interacting massive particles (FIMP) 
\cite{Hall:2009bx,Elahi:2014fsa,Biswas:2015sva,Biswas:2016bfo,
Bernal:2017kxu,Biswas:2019iqm,Barman:2020plp,Barman:2020ifq},
secluded sector DM \cite{Pospelov:2007mp,
Feng:2008mu,Chu:2011be,Berlin:2014pya,Foot:2014uba,Hambye:2019dwd} etc. 
For a recent review on DM production mechanisms beyond WIMP paradigm see
\cite{Baer:2014eja}. Among these
alternatives secluded sector DM is a promising scenario
to explain the negative results of the DM experimental
observations \cite{Pospelov:2007mp, Feng:2008mu, Foot:2014osa,Foot:2016wvj,Evans:2017kti}. In secluded sector DM, 
the dark sector contains the DM candidate and a metastable mediator
which couples with SM bath through a feeble portal coupling
and establishes the thermal equilibrium between dark and visible sector.
In this scenario, if the mass of the DM is greater than that of the mediator
then the relic density is governed by the annihilation cross section
of the DM into the mediator particles. Since the mediator can decay into
the SM particles therefore the portal coupling should be such that
it decays before the Big Bang Nucleosynthesis (BBN) so that  the
observations during BBN remains undistorted.

Due to the presence of the feeble portal coupling between the 
dark and visible sector, there is a possibility that the
dark sector may not be in thermal equilibrium with the SM bath.
In this case, the energy 
exchange between hidden and visible sector is still possible but
this exchange is not sufficient to equilibrate the two sectors. 
Therefore this non-adiabatic evolution of the dark sector 
opens up a new avenue of dark matter dynamics which is 
dubbed as \textit{``Leak In Dark Matter"} (LIDM) 
\cite{Evans:2019vxr}. In this scenario, since the dark
sector is not in thermal equilibrium with the visible sector therefore it has its 
own temperature provided the sector is internally thermalised.

Motivated by the production mechanisms for a gauged $B-L$ model discussed in \cite{Evans:2019vxr},
in this work, we have studied the non-adiabatic cosmological evolution
of the dark sector in $U(1)_{\lmt} \otimes U(1)_X$ extension of SM. Here, we have considered a 
dark sector which is invariant under $U(1)_X$ gauge
symmetry. The dark sector contains the DM candidate $\x$ which is a Dirac fermion,
and it is singlet under $\smgauge$ but charged
under $U(1)_X$. Now to connect the dark to the visible sector,
we have assumed that the SM is invariant under $U(1)_{\lmt}$
gauge symmetry \cite{He:1991qd,He:1990pn}.
The $U(1)_{\lmt}$ extension of SM
is very well motivated for various reasons. One of the 
interesting features of this model is that the extra gauge boson
corresponding to the $U(1)_{\lmt}$ can ameliorate the tension between SM 
prediction and experimental observation of muon $(g-2)$ 
\cite{PhysRevD.64.055006,Ma:2001md,Banerjee:2020zvi}. This model
has also been studied in the context of neutrino masses and mixing
in \cite{Ma:2001md}. Here, the tree-level kinetic mixing between $U(1)_X$ vector boson 
and the $U(1)_{\lmt}$ vector boson is present and 
the dark sector particles interacts with the visible sector
particles having non-zero $U(1)_{\lmt}$ charge.
Here we have assumed that the 
tree level kinetic mixing between $\lmt$ and $U(1)_Y$
vector bosons as well as $U(1)_X$ and  $U(1)_Y$ vector bosons are 
absent at tree level. Nevertheless, we will consider them to be generated
at the one loop level. We also assume that the $U(1)_X $ gauge boson $\zp$
becomes massive via St\"{u}ckelberg mechanism \cite{Stueckelberg:1938hvi,Ruegg:2003ps}.

Thus to study the different stages of the dark sector evolution, we have calculated 
the dark sector temperature $\tp$ by considering all the 
possible production channels which populate the dark sector radiation bath.
Considering a dark sector temperature $\tp$ which is not same as
the temperature of the SM bath ($T$), we 
have solved the Boltzmann equation numerically and depending 
upon the values of the model parameters such as portal coupling, 
DM mass, mass of the dark sector gauge boson, and dark 
sector gauge coupling we have studied
the different non-equilibrium states of the dark sector in detail.

Furthermore, the presence of $\lmt$ vector portal also
opens up the possibility of detecting the DM  via direct, indirect, 
laboratory, and astrophysical observations. Therefore we have studied 
the allowed model parameter space from various experiments.
In particular, we have investigated the allowed parameter space
from the direct detection experiments as well as cosmic
microwave background measurements by Planck. 
We have also studied the prospect of detecting $\gamma$-ray signal
from DM annihilation by considering a one step 
cascade process $\bar{\x}\x \ra \zp (\zp\ra \text{SM final states})$.
To constrain the model parameter space from the observation of diffuse $\gamma$-
ray background, we have calculated the $\gamma$-ray flux from DM annihilation
and compared with the measured diffuse $\gamma$-ray background flux by the
experimental collaborations such as
EGRET \cite{Strong:2004de},
COMPTEL \cite{comptel-kappadath}, INTEGRAL \cite{Bouchet:2011fn}, 
and Fermi-LAT \cite{Fermi-LAT:2012edv}.
The properties of the dark vector boson $\zp$ can be constrained
from various laboratory and astrophysical observations. Therefore
we have also studied the constraints on the mediator mass
$\zp$ and the portal coupling $\eps$ from BBN observations, beam-dump experiments, 
star and white dwarf cooling, SN1987A observations and fifth force searches.

Before ending this section, we would like to discuss a few things. 
The model studied in this work is different from
the gauged $B-L$ model, studied in \cite{Evans:2019vxr}, although different stages of dark sector
evolution are pretty much model independent. On the other hand, the detailed calculations
related to relic density are very much different in these two models. This is because
the SM quarks and first generation leptons do not couple with the dark vector 
boson at tree level. As a result of this, the limits on the parameter space of our model,
arising from direct, indirect, and astrophysical observations
are quite different in comparison to the gauged $B-L$ scenario \cite{Knapen:2017xzo, Evans:2019vxr}.
We have also included additional constraints such as muon $(g-2)$ and neutrino trident production at CCFR
in the present work.

The paper is structured as follows. In Section \ref{Model} we discuss 
our model briefly. We discuss the dynamics of the dark section in 
section \ref{Dynamics}. Section \ref{numerical_results} is devoted
to the numerical results for DM relic density. In section \ref{DM-indi} we discuss
direct detection, CMB constraint, and $\gamma$ ray signal
from DM annihilation. The constraints
on the $\lmt$ portal has been discussed in section \ref{con_zp} and finally
we summarise our results in section \ref{conclusion}.
A detailed calculation of gauge boson masses and 
their mixing angle for 
our model is presented in Appendix \ref{App:A}.
The calculation of the collision terms for temperature evolution
and calculation of reaction rates has been discussed in 
Appendices \ref{App:B} and \ref{App:C} respectively.
A brief discussion on the dark sector thermalisation
is given in Appendix \ref{App:D}.
Finally, the derivation of the photon spectrum for the final state radiation
is discussed in Appendix \ref{App:E}.

\section {The Model}
\label{Model}
In this section we discuss our model briefly. Here we have considered
a $U(1)_X \otimes U(1)_{\lmt}$ extension of the SM gauge group. 
The dark matter candidate $\x$ is a Dirac fermion and it is
singlet under $\smgauge\otimes U(1)_{\lmt}$ but charged
under the $U(1)_X$ gauge symmetry. Therefore the dark sector
contains the DM candidate $\x$ and a 
vector boson $\hat{Z}^\prime $ corresponding
to the $U(1)_X$ gauge group. Additionally we also consider
the $U(1)_{\lmt}$ gauge symmetry in the SM Lagrangian
and the corresponding vector boson is denoted by $\hat{Z}_{\mt}$.
Since $U(1)_{\lmt}$ is an anomaly free gauge theory therefore
we do not need any extra chiral fermions to cancel the gauge
anomaly. Thus the tree-level kinetic-mixing between
$U(1)_X$ and $U(1)_{\lmt}$ gauge bosons establishes the 
connection between the dark and the visible sector.
As already mentioned in the Introduction, the tree-level
kinetic mixing between $U(1)_Y$ gauge boson $B$ and 
$\hat{Z}_{\mt}$ as well as $B$ and $\hat{Z}^\prime$ are absent.
However, these kinetic mixings can be generated
radiatively and the impact of these radiatively generated
kinetic mixings will be discussed later.

Thus the Lagrangian for our model is given by
\bea
\label{Lagrangian}
\mathcal{L}
&=&
\mathcal{L}_{\rm SM}  + \bar{\x}\left(i \slashed{\partial} - m_\x\right)\x
- \dfrac{1}{4} \hat {X}^{\rho \sigma} \hat{X}_{\rho \sigma}
- \dfrac{1}{4} \hat {F}_{\mt}^{\rho \sigma} \hat{F}_{\mt_{\rho \sigma}}
\nn\\
&&
- g_X \bar{\x} \gamma^\rho \x {\hat{\zp}}_\rho
-g_{\mu \tau}
\left(
\bar{\mu}\gamma_\rho \mu + \bar{\nu}_\mu \gamma_\rho P_L \nu_\mu
-\bar{\tau}\gamma_\rho \tau - \bar{\nu}_\tau \gamma_\rho P_L \nu_\tau
\right)\hat{Z}_{\mt}^\rho\nn\\
&&
+\dfrac{1}{2} \hat{m}_{\mt}^2 \hat{Z}_{\mt}^\rho 
\hat{Z}_{\mt_\rho}
+\dfrac{1}{2} \hat{m}^{\prime 2} \hat{\zp}^\rho \hat{\zp}_\rho
+ \dfrac{\sin \delta}{2} \hat{F}_{\mu\tau}^{\rho \sigma} \hat{X}_{\rho \sigma}\,\,.
\eea
Here $g_{\mt}$ and $g_X$ are the gauge couplings of the additional
$U(1)_{\lmt}$ and $U(1)_X$ gauge symmetry respectively. Similarly
$\hat{X}_{\rho \sigma} = \partial_\rho \hat{\zp}_\sigma - 
\partial_\sigma \hat{\zp}_\rho$ and 
$\hat{F}_{\mt}^{\rho \sigma} = \partial^\rho \hat{Z}_{\mt}^\sigma - 
\partial^\sigma \hat{Z}_{\mt}^\rho$
are the field strength tensor of $U(1)_X$ and $U(1)_{\lmt}$ 
gauge symmetry respectively. $m_\x$ is the mass of the DM 
and since the origin of the mass parameters $\hat{m}_{\mt}$ 
and $\hat{m}^\prime$ are not relevant for our work therefore we have considered that they 
are generated via St\"{u}ckelberg mechanism \cite{Stueckelberg:1938hvi,Ruegg:2003ps}.
The last term of Eq. \ref{Lagrangian} indicates the kinetic mixing 
between $\hat{Z}^\prime$ and $\hat{Z}_{\mt}$.

To proceed further, we need to express Eq. \ref{Lagrangian} in canonical form. 
To do this, we have performed a non orthogonal transformation from
the ``\textit{hatted}" basis to an ``\textit{barred}" basis i.e.
from $\left(\hat{Z}_{\mt_\rho} ~~~ \hat{Z}^\prime_\rho \right)^T$ 
to $\left(\bar{Z}_{\mt_\rho} ~~~ \bar{Z}^\prime_\rho \right)^T$
to remove the last term of Eq. \ref{Lagrangian}. 
The non orthogonal transformation is given by 
\bea
\label{non-ortho-trans_1}
\begin{pmatrix}
\hat{Z}^\prime_\rho \\
\hat{Z}_{\mt_\rho}
\end{pmatrix}
 =
\begin{pmatrix}
\sec \delta & 0\\
\tan \delta & 1
\end{pmatrix}
\begin{pmatrix}
\bar{\zp}_\rho\\
\bar{Z}_{\mt_\rho}
\end{pmatrix}
\eea
However, the removal of the kinetic mixing term generates
a mass mixing between $\bar{Z}^\prime$ and $\bar{Z}_{\mt}$. 
Therefore the gauge boson mass matrix in 
$\left(\bar{Z}_{\mt_\rho} ~~~ \bar{Z}^\prime_\rho \right)^T$
basis is given by the following $2\times$2  symmetric
matrix.
\bea
\label{mass-matrix}
\mathcal{M}_{\rm GB}^2
&=&
\hat{m}_{\mt}^2
\begin{pmatrix}
1&& \tan \delta\\
\tan\delta && \tan^2 \delta + \dfrac{\kappa^2}{\cos^2 \delta}
\end{pmatrix}\,\,,
\eea
where $\kappa^2 = \hat{m}^{\prime 2}/ \hat{m}_{\mt}^2 $. To diagonalise
the above mentioned mass matrix we perform
an orthogonal rotation in the 
$\left(\bar{Z}_{\mt_\rho} ~~~ \bar{Z}^\prime_\rho \right)^T$
plane to diagonalise the gauge boson mass matrix.
The orthogonal transformation is given by
\bea
\label{ortho-trans}
\begin{pmatrix}
\bar{Z}_{\mt_\rho}\\
\bar{Z}^\prime_\rho
\end{pmatrix}
&=&
\begin{pmatrix}
\cos \beta & -\sin \beta\\
\sin \beta & \cos \beta
\end{pmatrix}
\begin{pmatrix}
Z_{\mt_\rho}\\
\zp_\rho
\end{pmatrix}
\,\,\,,
\eea
where $\beta$ is the mixing angle between $\zp$ and $\zmt$.
Therefore, we can express Eq. \ref{Lagrangian} in the mass basis
of the gauge bosons by using Eq. \ref{non-ortho-trans_1} and
Eq. \ref{ortho-trans}. In this work we have
assumed $m_{\mt}^2>> m_{\zp}^2$ where $m_{\mt}$ and 
$m_{\zp}$ are masses of $Z_{\mt}$ and $\zp$ respectively.
Using this assumption, 
we can write the interaction of the DM
with $\zp$ and the portal interaction in the mass basis
of the gauge bosons as
\bea
\mathcal{L}
\label{Lagrangian_2}
\supset
-g_X \bar{\x} \gamma^\rho \x \zp_\rho
+\eps\left(
\bar{\mu}\gamma^\rho \mu + \bar{\nu}_\mu \gamma^\rho P_L \nu_\mu
-\bar{\tau}\gamma^\rho \tau - \bar{\nu}_\tau \gamma^\rho P_L \nu_\tau
\right)\zp_\rho\,\,\,.
\eea
In the above $\eps = g_{\mt} \hat{\eps}$ is the portal coupling between 
dark and the visible sector and it depends on 
$U(1)_{\lmt}$ gauge coupling $g_{\mt}$, 
kinetic mixing parameter
$\sin \delta$, $m_{\zp}$, and $m_{\mt} $.
In Appendix \ref{App:A} we have provided the calculation
of the parameter $\hat{\eps}$. 

\section {Dark sector dynamics}
\label{Dynamics}
As discussed in the previous section, our model
has four free parameters such as 
$m_\x$, $m_{\zp}$, $g_X \text{ or } \alpha_X = g_X^2/4 \pi$, and $\eps$.
Furthermore, we have considered $\zmt$ is in thermal equilibrium with other
SM species and $m_{\mt}$ is much heavier than $m_{\zp}$. Therefore, $\zmt$
does not play any role in the dynamics of the dark sector. 
In this work we are interested to explore the 
time evolution of a secluded dark sector which is
not in thermal contact with the SM bath.
Since the two sectors might achieve thermal equilibrium
through the portal coupling $\eps$, hence we choose
$\eps$ to be much smaller than 1 for thermally decoupled
dark sector. As the dark sector is not in thermal equilibrium
with the SM bath, it may have its own temperature $\tp$
provided the dark sector is internally thermalised and in that case
the evolution of $\tp$ is completely different in comparison
to the SM bath temperature $T$. The different time
evolution of $\tp$ have great impact on the DM relic density
and the model parameter space changes significantly compared
to the standard WIMP scenario.

Let us note in passing that throughout the paper all
the dark sector quantities are denoted with a \textit{prime}
whereas for the visible sector the corresponding quantities
are denoted \textit{without a prime}.

To start with first let us discuss the temperature
evolution of the hidden sector thermal bath.

\subsection{Dark sector temperature evolution}
\label{DS_temp}
The temperature of the dark radiation bath 
depends on the portal coupling between the
hidden and the visible sector. 
For large portal coupling it is possible that
the two sectors are in thermal equilibrium and they have
a common temperature.
However for small values of the portal coupling,
the dark sector is not in thermal equilibrium with 
the SM bath and out of equilibrium energy injection 
from visible to dark sector
is still possible. It is known as the \textit{non-adiabatic}
evolution of the dark sector and in this case the
dark sector temperature $\tp$ 
evolves in an unconventional
manner.

To study the evolution of $\tp$, we need to solve
the Boltzmann equation (BE) for the energy density of the dark sector $\rp$.
The time evolution of $\rp$ is given by the following BE.
\bea
\label{BE_rho}
\dfrac{d \rp}{d t} + 4 H \rp = \mathcal{C}_{{\rm SM} \rightleftarrows \zp} (T,\tp)\,\,\,.
\eea
In the above, $H$ is the Hubble parameter which is given by
\bea
\label{Hubble}
H &=& \sqrt{\dfrac{8\pi}{3 M_{\rm Pl^2}} \left(\rho_{\rm SM} + \rp\right)}\,\,\,,
\eea
where $M_{\rm Pl} = 1.22\times 10^{19}$ GeV is the Planck mass 
and $\rho_{\rm SM} = \dfrac{\pi^2}{30} g_{\rho}(T) T^4$ 
is the energy density of the SM bath where $g_{\rho} (T)$ is the relativistic
degrees of freedom contributing to the SM energy density. In the right hand side (r.h.s)
of Eq. \ref{BE_rho}, $\mathcal{C}_{{\rm SM} \rightleftarrows \zp} (T,\tp)$ is the relevant
collision term for energy exchange between visible and dark sector.
For a process like $SM (P_1) + SM (P_2) \rightarrow SM(P_3) + \zp (P_4)$
the explicit form of the collision term is given by
\bea
\label{Coll_term}
\mathcal{C}_{{\rm SM}\rightleftarrows \zp} (T, \tp)
&=&
\sum_{\rm {All\, channels}} \int d \Pi_i E_4
(2 \pi)^4 \delta^4 (P_1 + P_2 - P_3 -P_4)
\overline{|\mathcal{M}|^2}
f_{\rm SM} (p_1, T) f_{\rm SM}(p_2, T) \nn\\
&-&
\sum_{\rm {All\, channels}} \int d \Pi_i E_4
(2 \pi)^4 \delta^4 (P_1 + P_2 - P_3 -P_4)
\overline{|\mathcal{M}|^2}
f_{\rm SM}(p_3, T) f_{\zp} (p_4, \tp)\nn\\
&=&
\mathcal{C}_{{\rm SM}\to \zp}(T) - \mathcal{C}_{\zp \to {\rm SM}}(T,\tp)\,\,,
\eea
where $d \Pi_i = g_i\dfrac{d^3 \vec{p}_i}{(2 \pi)^3 2 E_i}$ is the Lorentz invariant
phase space measure, $g_i$ is the internal degrees of freedom of the $i^{th}$ species,
$\overline{|\mathcal{M}|^2}$ is the matrix amplitude square averaged
over initial and final states, $P_i$ and $m_i$ are 
the four momentum and mass of the $i^{th}$ species respectively,
and $p_i = |\vec{p}_i|$. In the r.h.s. of Eq. \ref{Coll_term} the first
term is the collision term for the energy injection from visible to dark sector
whereas the second term is the collision term for the energy injection
from dark to visible sector.

Now using the technique discussed in \cite{Gondolo:1990dk}, we can write
the collision term $\mathcal{C}_{{\rm SM} \to \zp}(T)$
as follows:
\bea
\label{coll_term_final_text}
\mathcal{C}_{{\rm SM}\to \zp} (T)
&=&
\dfrac{g_3 g_4 T}{ (2 \pi)^4}
\int_\Upsilon^\infty
\lambda^2\,\hat{s}^2\,\left(1 - \eta \right)
\sigma_{34\to 12} \, K_2 \left(\dfrac{\sqrt{\hat{s}}}{T}\right)
d \,\hat{s}\,\,\,\,,
\eea
where $\sigma_{34 \to 12}$ is the annihilation cross section of 
$SM (P_3) + \zp (P_4) \rightarrow SM(P_1) + {\rm SM} (P_2)$ process,
$\hat{s} = (P_3 + P_4)^2$ is the Mandelstam variable, 
$K_2(\sqrt{\hat{s}}/T)$ is the Bessel function of second kind and order two.
The other quantities such as $\Upsilon$, $\lambda$, and $\eta$ are defined
as follows:
\bea
\Upsilon &=& {{\rm Max}\left[(m_1 + m_2)^2, (m_3 + m_{\zp})^2\right]}\,\,,\nn\\
\lambda &=& \dfrac{\sqrt{\hat{s} - (m_3 + m_4)^2}
\sqrt{\hat{s}- (m_3^2 - m_4)^2}}{2 \hat{s}}\,\,,\nn\\
\eta&=& \dfrac{m_3^2 - m_4^2}{\hat{s}}\,\,,
\eea
where $m_{\zp}$ is the mass of the $\zp$ gauge boson.
A detailed discussion on the calculation of the collision
term $\mathcal{C}_{{\rm SM}\to \zp} (T)$ is given 
in Appendix \ref{App:B}.

\begin{figure}
\centering
\includegraphics[height=7cm,width=9cm]{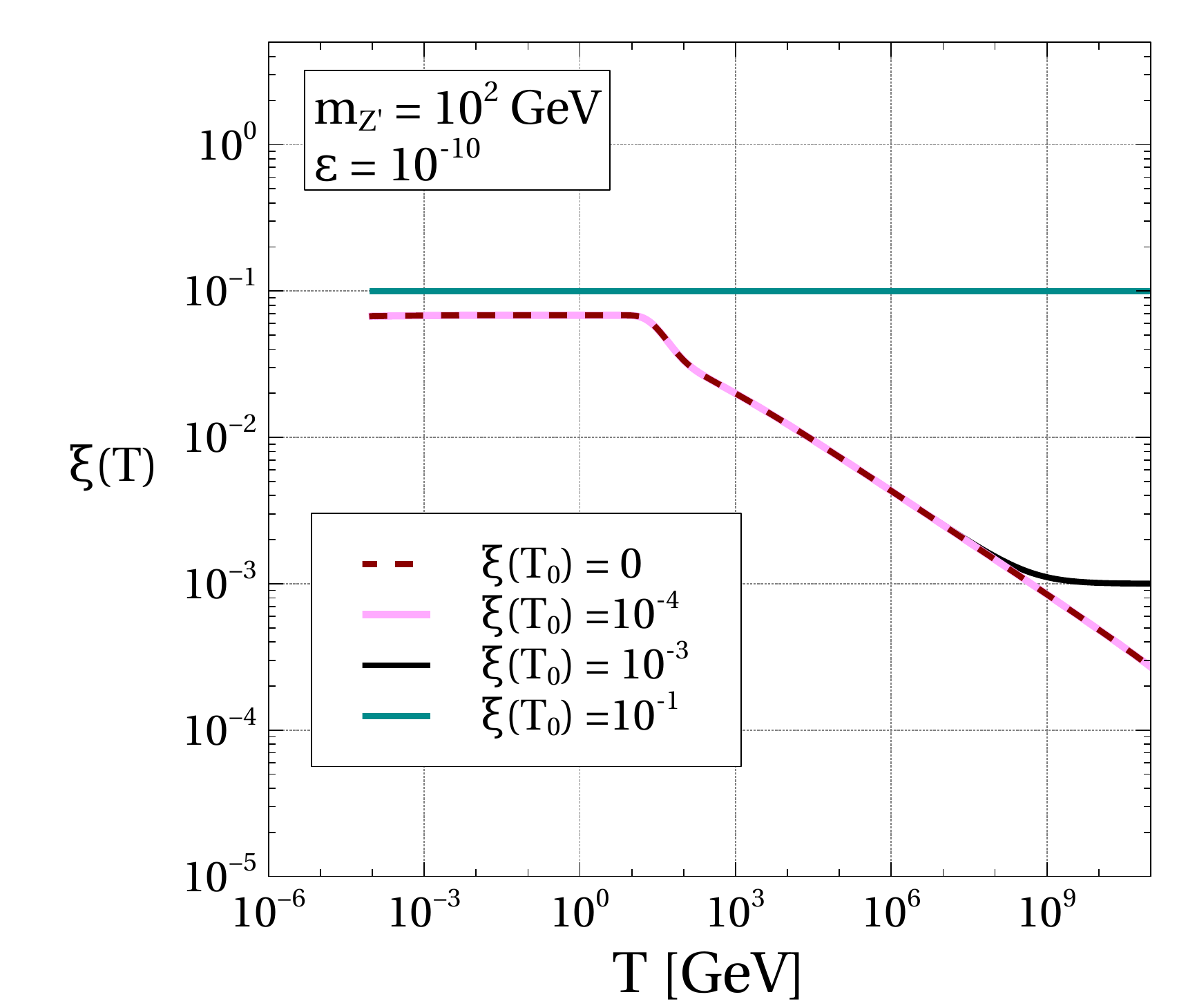}
\caption{Evolution of $\xi (T)$ as a function of $T$ for
$\xi(T_0) = 0$ (dotted brown), $10^{-4}$ (solid pink), $10^{-3}$ (solid black),
$10^{-1}$ (solid green). Here we choose $m_{\zp} = 10^2\,\rm GeV$ and $\eps = 10^{-10}$.}
\label{fig:XI_vs_T_diff_XI0}
\end{figure}

In order to solve Eq. \ref{BE_rho} we have assumed the following conditions.
\begin{itemize}
\item The dark sector is internally thermalised 
(see Appendix \ref{App:D} for details) and therefore
we can write $\rp = \dfrac{\pi^2}{30} g_{\rp} {\tp}^4$ where
$g_{\rp}$ is the relativistic degrees of freedom contributing to the radiation
bath of the dark sector and it is taken to be 3 throughout our analysis. This is because
in the region of our interest $\zp$ is always relativistic.

\item The energy density of the Universe is dominated by the SM bath. This
is because the dark sector is colder than the visible sector i.e. $T>>\tp$.

\item The dark sector is not in thermal contact with the SM bath
and we have neglected the energy injection from dark to visible sector i.e. we have 
taken $\mathcal{C}_{\zp \to {\rm SM}} (T, \tp) \simeq 0$ throughout our analysis.

\item The entropy density of the SM sector is approximately conserved.
\end{itemize}
Using the above assumptions and defining $\xi = \tp/T$,
we can express Eq. \ref{BE_rho} as 
\bea
\label{xi_inter_1}
4 \xi^3 \dfrac{d \xi}{d T}
&\simeq& - \dfrac{30\, \mathcal{C}_{{\rm SM}\to \zp} (T)}{g_{\rp} \pi^2 T^5 H(T)}\,\,.
\eea
Therefore the final form of $\xi(T)$ as a function of $T$ can be obtained
from Eq. \ref{xi_inter_1} and it is given by
\bea
\label{xi-eq}
\xi (T)
&=&
\left[\int_{T}^{T_0}
\dfrac{30 \,\mathcal{C}_{{\rm SM}\to \zp} (\tilde{T})}
{g_{\rp} \pi^2 H(\tilde{T}) \tilde{T}^5}
d \tilde{T}\right]^{1/4}\,\,\,,
\eea
where $T_0$ is the initial temperature of the early
Universe and we take $\tp \simeq 0$ as $T = T_0$.
In our scenario, the dark sector is produced
from the SM bath and we choose the initial value of the $\tp$
to be zero. In Fig.\,\ref{fig:XI_vs_T_diff_XI0}, we show the variation of the quantity $\xi(T)$
with the SM bath temperature $T$ for different choices of $\xi (T_0)$.
One can see from this figure that for $\xi(T_0) \lesssim 10^{-4}$, the dark sector
temperature is insensitive to the choice of $\xi (T_0)$. In principle,
it is always possible to have a temperature asymmetry between the visible and dark
sector in the early Universe and one such example is the production of SM and
dark sector particles from the decay of inflaton. However, the study of 
the dynamics of inflaton decay and the energy exchange between visible and dark sector
is quite involved and it is beyond the scope of this paper.

The $T$ dependence of $\xi$ relies on the fact
whether the portal interactions are  renormalisable or non-renormalisable.
For a renormalisable interaction, the annihilation cross section is
$\sim$ $1/\hat{s}$ where $\hat{s}$ is the Mandelstam variable.
If we plug this in Eq. \ref{coll_term_final_text} then 
$\mathcal{C}_{{\rm SM}\to\zp} (T) \sim T^{5}$ 
and using this collision term we can calculate
the $T$ dependence of $\xi$ which is
given by $\xi (T) \sim T^{-1/4}$ (assuming $T_0 \gg T$).
However the dependence is drastically different in case 
of non-renormalisable portal interaction. For a
non-renormalisable portal interaction we can write
the annihilation cross section $\sim$ $1/\Lambda^2$
where $\Lambda$ is the cut-off scale of the theory.
In that case the collision term $\mathcal{C}_{{\rm SM}\to\zp}$
goes as $T^7/\Lambda^2$ and $\xi(T) \propto (T_0^{1/4} - T^{1/4})$.
If we assume that $T_0$ to be the maximum temperature of
the Universe i.e. reheat temperature $T_{\rm RH}$ then
$\xi(T)$ is proportional to $T_{\rm RH}^{1/4}$.
Let us note that, in this calculation we have assumed $\Lambda > T_{\rm RH}$.
Therefore, for renormalisable portal interaction,
$\xi(T)$ is sensitive to the SM temperature $T$
and it increases with the decrease in $T$ i.e.
the physics for renormalisable portal interaction
is effective in the low energy scale. 
However, for non-renormalisable portal 
interaction, $\xi(T)$ is sensitive to the
early history of the Universe i.e. the
value of $T_{\rm RH}$. 

To calculate the dark sector temperature $\tp$ we have considered
all the processes\footnote{Here we have not considered $\bar{f}f \to \zp \zp$
since at the amplitude level  these processes are proportional to $\eps^2$.}
which produce the dark vector boson $\zp$.
The relevant processes can be classified into two categories:

\vspace{5pt}
\noi
a) Processes involving neutral current interactions such as
$f \gamma \to   f \zp$, 
$f  Z \to  f \zp$, 
$f h \to f \zp$, 
$\bar{f} \gamma \to \bar{f} \zp$,
$\bar{f} Z \to \bar{f} \zp$,
$\bar{f} h \to \bar{f} \zp$,
$\bar{f}  f\to \gamma  \zp$,
$\bar{f} f\to Z \zp$, 
$\bar{f}f \to h \zp$, 
$\bar{f}{f} \to \zp $
where
$f = \{\mu, \tau, \nu_\mu, \nu_\tau\}$, $Z$ and $h$ are the 
SM $Z$ boson and SM Higgs boson respectively, 

\vspace{5pt}
\noi
b) Processes involving charged current interactions such as
$l \, W^+ \to \nu_l \, \zp$, 
$\nu_l \, W^- \to l\, \zp$,
$\bar{l} W^- \, \to \bar{\nu_l}\, \zp$,
$\bar{\nu}_l\,  W^+ \to \bar{l}\,  \zp$,
$\bar{l}\, \nu_l \to W^+ \zp$,
$\bar{\nu}_l \, l \to W^- \zp$ where
$l = \{\mu, \tau\}$, $W^\pm$ is the $W$ gauge boson of SM.

\begin{figure}
\centering
\includegraphics[height = 7cm, width = 8cm]{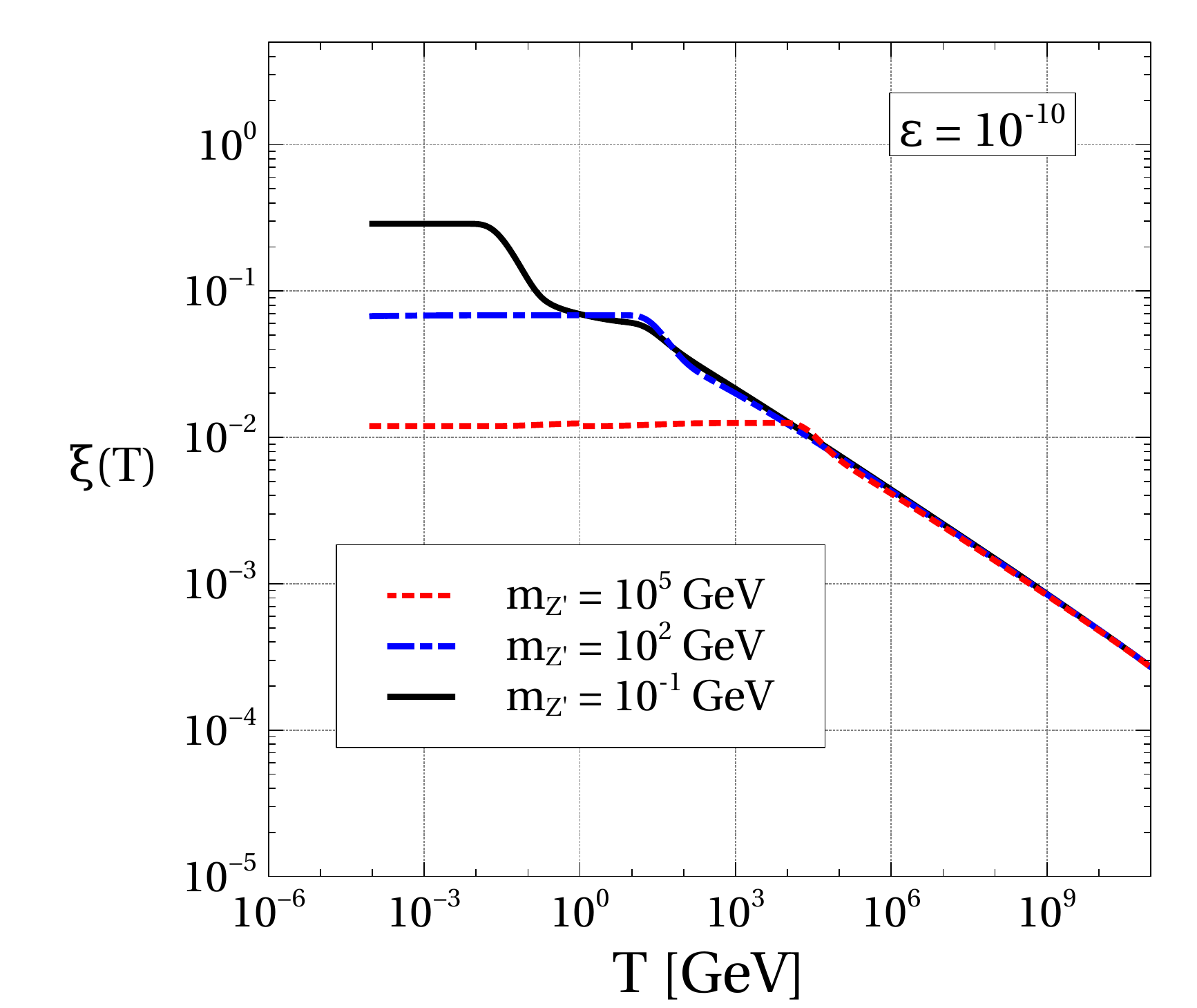}
\includegraphics[height = 7cm, width = 8cm]{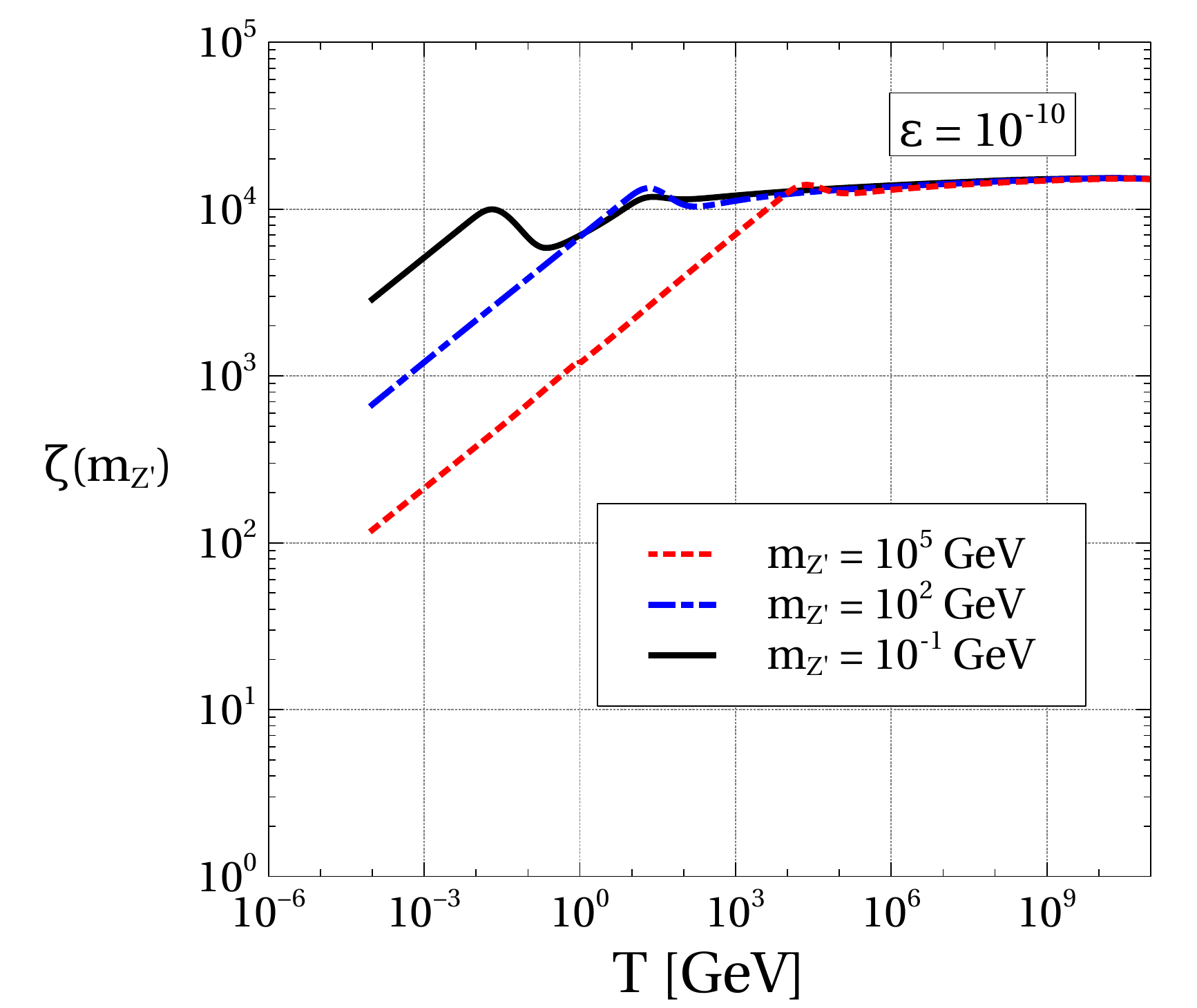}
\caption{
{\textbf{\textit{Left panel}}}:
Variation of $\xi(T)$ as a function of $T$ for 
$m_{\zp} = 10^{-1} \, {\rm GeV}$ (solid black line), 
$m_{\zp} = 10^{2} \, {\rm GeV}$ (blue dashed-dot line),
$m_{\zp} = 10^{5} \, {\rm GeV}$ (red dotted line).
{\textbf{\textit{Right panel}}}: Variation of $\zeta (m_{\zp})$
as a function of $T$ and the values of $m_{\zp}$ and color codes
are same as left panel. In both the plots we have considered
$\eps = 10^{-10}$.
}
\label{xi_vs_T}
\end{figure}

Considering all the processes mentioned above we solve Eq. \ref{xi-eq}
numerically and the numerical results are presented in 
the left panel of Fig.\,\ref{xi_vs_T}.
We can see from the figure that for large value of $T$, $\xi(T)$
increases with the decrease in $T$ and it remains constant
below $T\lesssim m_{\zp}$ i.e. when all the 
energy injection processes stop. Since maximum
production from inverse decay processes will occur
at $T\sim m_{\zp}$ therefore there is a small kink
at $T\simeq m_{\zp}$ in the figure for all
values of $m_{\zp}$.

Since the collision term for all the production channels of $\zp$
are proportional to $\eps^2$ therefore $\xi (T)$ should be proportional
to $\sqrt{\eps}$. Thus to get a semi-analytic expression of $\xi(T)$, we 
parameterize $\xi(T)$ in the following manner.
\bea
\xi(T) &=& \zeta (m_{\zp}) \,\sqrt{\eps} \,T^{-1/4}\,\,,
\eea
where $\zeta (m_{\zp})$ is a function of $m_\zp$ and
the choice of the temperature dependence is motivated
by the discussion below Eq. \ref{xi-eq}. To determine
the behaviour of $\zeta(m_\zp)$ we have calculated
$\zeta(m_{\zp})$ from our numerical results and plotted
as a function of $T$ for three different values of 
$m_{\zp}$ and $\eps = 10^{-10}$ in the right panel of 
Fig.\,\ref{xi_vs_T}. 
From the figure we can see that 
$\zeta(m_\zp)$ does not depend on $m_{\zp}$ and $T$
for $T\gtrsim \, 10 m_{\zp}$ and in this region the 
value of $\zeta(m_\zp) \simeq 10^4$
Therefore we can write the semi-analytic form of $\xi (T)$ for 
$T>10\,m_{\zp}$ as
\bea
\label{xi-approx}
\xi(T) \simeq 10^4 \sqrt{\eps} \, T^{-1/4}\,\,.
\eea
Here, we would like to mention that, in our scenario, dark and visible
sectors are not in thermal equilibrium and the value of 
$\xi$ depends on the portal coupling $\eps$.

\subsection{Non-adiabatic states of the dark sector}
In this section we discuss various stages the
of the dark sector during its cosmological
evolution. Depending on the parameters of the dark sector,
such as $m_\x$, $m_{\zp}$, $\alpha_X$ and also
the portal coupling $\eps$, the states
of the dark sector can be broadly classified into 
two parts and they are $i)$ equilibrium state
and $ii)$ non-equilibrium state.

\subsubsection{Equilibrium state}
\label{wimp-next-door}
In case of non-adiabatic evolution, if the portal coupling 
is sufficiently high then the two sector equilibrate
and they share a common temperature. 
This framework resembles with the initial
idea of secluded sector DM in which
the DM freezes out of a dark radiation bath 
which is in thermal equilibrium
with the visible sector. Recently
this scenario is named as ``WIMP next door"
\cite{Evans:2017kti}.

In this work, we are interested in the scenario
in which the dark sector is not in thermal equilibrium with
the visible sector. To identify the parameter space
in the  $m_\x-\eps$ plane
for the thermally decoupled dark sector,
first we identify the allowed parameter space in $m_\x-\eps$ plane
for the ``WIMP next door" scenario. The disallowed region
for the ``WIMP next door" is the region of interest
of our analysis.

For that we have assumed $\x$, $\zp$ are in thermal equilibrium
with the SM bath. We have calculated
total reaction rate ($\Gamma$)
for the processes mentioned in section
\ref{DS_temp} and compared with the Hubble parameter
$H(T)$. The ratio $\Gamma/H > 1$ at $T\simeq m_\x/20$
gives the allowed region for the ``WIMP next door".
Note that we have considered the contribution
of $\zp$ in the Hubble parameter since it is assumed that
$\zp$ is in thermal equilibrium with the SM bath.
A detailed discussion on the calculation of $\Gamma$
is given in Appendix \ref{App:C}.

\subsubsection{Non-equilibrium state}
\label{non-eq-phase}
If the dark sector is not in thermal equilibrium
then it may have its own temperature as we mentioned earlier.
In this case depending on the model parameters such as $\eps$, $\alpha_X$, $m_\x$
there are three possible stages of dark sector evolution and they are 
i) leak in, ii) freeze-in, iii) reannihilation.
\paragraph{Leak in:}
The freeze-out of DM from the dark radiation bath
during the production of dark radiation i.e. $\zp$
from the SM bath is known as `Leak in dark matter' (LIDM).
In this scenario the final abundance of the 
DM depends on the ratio $\xi$ which is defined
as $\xi = \dfrac{\tp}{T}$ and it is a function
of $T$.

Nevertheless, there are clear differences between the
dark sector freeze-out and leak in scenario. In the case of dark sector
freeze-out it is assumed that the dark sector evolves adiabatically
and the dark sector temperature $\tp = \xi T$, where $\xi$
is a constant quantity. In case of LIDM, the scenario is a little
different. In this case the dark sector evolves non-adiabatically
and DM freezes out during this non-adiabatic evolution.

To understand the LIDM mechanism quantitatively, let
us discuss the abundance of DM in the leak in scenario.
As discussed in section \ref{DS_temp}, the dark sector
temperature during the energy injection epoch
can be written as $\tp = \xi T$ 
where $\xi$ is a function of $T$. Now to get
an approximate estimate of DM abundance we use
sudden freeze-out condition $n_{\x}(\tp_f) \langle \sigma v\rangle
\simeq H (T_f)$ where $T_f(\tp_f)$ is the temperature
of the visible (dark) sector at the time of DM freeze-out,
$\langle \sigma v \rangle$ is $2\to2$ DM annihilation cross section, 
and $n_\x (\tp_f)$ is the DM number density 
at the time of DM freeze-out.
Using the sudden freeze-out condition, the relic density of DM
is given by
\bea
\label{relic_approx}
\Omega h^2
&\simeq &
0.12 \left( \dfrac{\xi}{10^{-5}}\right)
\left(\dfrac{x_f^\prime}{10}\right)
\left(\dfrac{7\times 10^{-15}\,{\rm GeV}^{-2}}{\langle \sigma v \rangle}\right)
\left(\dfrac{\sqrt{g_\rho (T_f)}}{10}\right)
\left(\dfrac{100}{g_{*s} (T_f)}\right)\,\,\,,
\eea
where $g_\rho (T)$ and $g_{*s} (T)$ are the relativistic
degrees of freedom contributing to the energy
density and the entropy density of the SM respectively
and $x^\prime_f = m_\x/T^\prime_f$.

Now we can see from Eq. \ref{relic_approx} that the required
cross section to get correct relic abundance is much smaller than
the required cross section ($\approx$ $10^{-8} \, {\rm GeV}^{-2}$) 
for standard WIMP scenario. 
Thus LIDM scenario naturally indicates
that the mass of the DM should be heavy and the relevant
discussion is given below.

The upper limit of the DM mass can be set
from the S-matrix unitarity \cite{Griest:1989wd}
in the LIDM scenario. The s-wave term of the thermally averaged
DM annihilation cross section is given by \cite{Griest:1989wd}
\bea
\langle \sigma v \rangle_{\rm Max}
&\simeq& \dfrac{4 \pi}{m_\x^2}\sqrt{\dfrac{x^\prime_f}{\pi}}\,\,.
\label{cross-uni}
\eea
Thus using Eq. \ref{cross-uni} in Eq. \ref{relic_approx} we get 
the upper bound of the DM mass which is given by
\bea
m_\x \lesssim \dfrac{127 \, \rm TeV}{\sqrt{\xi}}\,\,.
\eea
Thus $\xi$ plays an important role in determining the upper limit of $m_\x$
and for $\xi$ to be much smaller than 1, the upper
limit of the DM mass is much larger in comparison
to the standard WIMP scenario. For example
if we consider $\xi = 10^{-5}$ and $\x$ to be dirac fermion 
then $m_\x \lesssim 40 \,{\rm PeV}$
whereas for standard WIMP scenario the upper
limit of the DM mass is $m_\x \lesssim 127 \, \rm TeV$.
In deriving these numbers we have used the sudden freeze-out
approximation and $x^\prime_f \simeq 10$.

\paragraph{Freeze-in:}
Feebly interacting massive particle (FIMP)
is a well studied scenario and it is motivated
by the null results of (in)direct searches.
In this scenario the DM is produced out of equilibrium
from the SM bath and the production stops
at $T\sim {\rm Max}[m_\x, m_{\rm SM}]$ where
$m_{\rm SM}$ is the mass of the parent particles. Therefore 
the abundance of the DM is set by out of equilibrium
production of DM from SM bath.
Since the DM is out of equilibrium with the SM bath therefore
its coupling with the SM is very weak ($\sim 10^{-10}$)
and it can easily explain the null results of the experimental
searches.

In our model, the DM particle $\x$ 
is in thermal equilibrium with the dark radiation
bath and it follows the equilibrium number density.
After decoupling from the dark radiation bath,
the abundance of $\x$ can be increased due to the
presence of the DM production channel
$\bar{f} f \to \bar{\x} \x$ 
where ($f = \mu, \tau, \nu_\mu, \nu_\tau$).
Since we have assumed $\eps<<1$ therefore 
the abundance of $\x$ freezes at $T \sim {\rm Max}[m_f, m_\x]$
provided $\alpha_X$ is not sufficiently high. 
Thus in our scenario it is possible that the final
DM abundance is set by the freeze-in mechanism.

\paragraph{Reannihilation:}
If the coupling between the dark sector particles
are sufficiently high then there is a
possibility that the excess DM produced via freeze-in
can reannihilate into the dark sector particles.
In this situation if the rate of production of DM via freeze-in and the rate
of depletion via reannihilation are equal then it
follows a quasi static equilibrium and finally it freezes-out when the
DM decouples from the quasi-static equilibrium
\cite{Cheung:2010gj,Chu:2011be}. This phenomenon
is known as reannihilation.
In our model, for larger values of $\alpha_X$ compared to the
freeze-in scenario, it is possible that the DM produced
via freeze-in, reannihilate into the dark vector boson $\zp$.
Therefore in this case final abundance of the DM is set
by the reannihilation mechanism.

\subsection{Boltzmann Equation}
In this section we formulate the Boltzmann equation
for the evolution of the DM number density. As discussed
in section \ref{BE_rho} that
the dark sector temperature
can be expressed as a function of $T$ and $\eps$. 
Therefore we can obtain the final relic abundance of DM
by solving the Boltzmann equation considering
a different dark sector temperature ($\tp$). The Boltzmann
equation for the number density of $\x$ is given
by
\bea
\dfrac{d n _{\x_{\rm tot}}}{d t} + 3 H n_{\x_{\rm tot}}
&=& \dfrac{1}{2}\left[
\langle \sigma v\rangle ^{\tp}_{\bar{\x}\x\to \zp \zp}
\left(n_{\x_{\rm eq_{\rm tot}}} (\tp)^2 - n_{\x_{\rm tot}}^2\right)
\right] \nn\\
&+&
2 \sum_f \langle \sigma v\rangle^T_{\bar{f}{f}\to \bar{\x}{\x}}
n_{f_{{\rm eq}}}^2(T)
\label{BE_n}
\eea
The first term on the left hand side  of Eq. \ref{BE_n} denotes
the change in DM number density ($n_{\x_{\rm tot}} = n_\x + n_{\bar{\x}}$) whereas
the second term implies the dilution of $n_{\x_{\rm tot}}$
due to expansion of the Universe. The first term
on the right hand side of Eq. \ref{BE_n} denotes
the DM interaction with the $\zp$ bath 
and $\langle \sigma v \rangle^{\tp}_{\bar{\x}{\x} \to \zp \zp}$
is thermally averaged cross section of $\bar{\x}{\x}\to \zp\zp$
calculated at temperature $\tp$. The s-wave term of the thermal average
of the $\bar{\x}{\x} \to \zp \zp$ cross section is given by
\bea
\label{cross-swave}
\langle \sigma v \rangle_{\bar{\x}{\x} \to \zp \zp}
&\simeq& \dfrac{4 \pi \alpha_X^2}{m_\x^2}
\dfrac{m_\x\left(m_\x^2 - m_{\zp}^2\right)^{3/2}}
{\left(2 m_\x^2 - m_{\zp}^2\right)^2}\,\,.
\eea

\begin{figure}
\centering
\includegraphics[height = 9cm, width = 12cm]{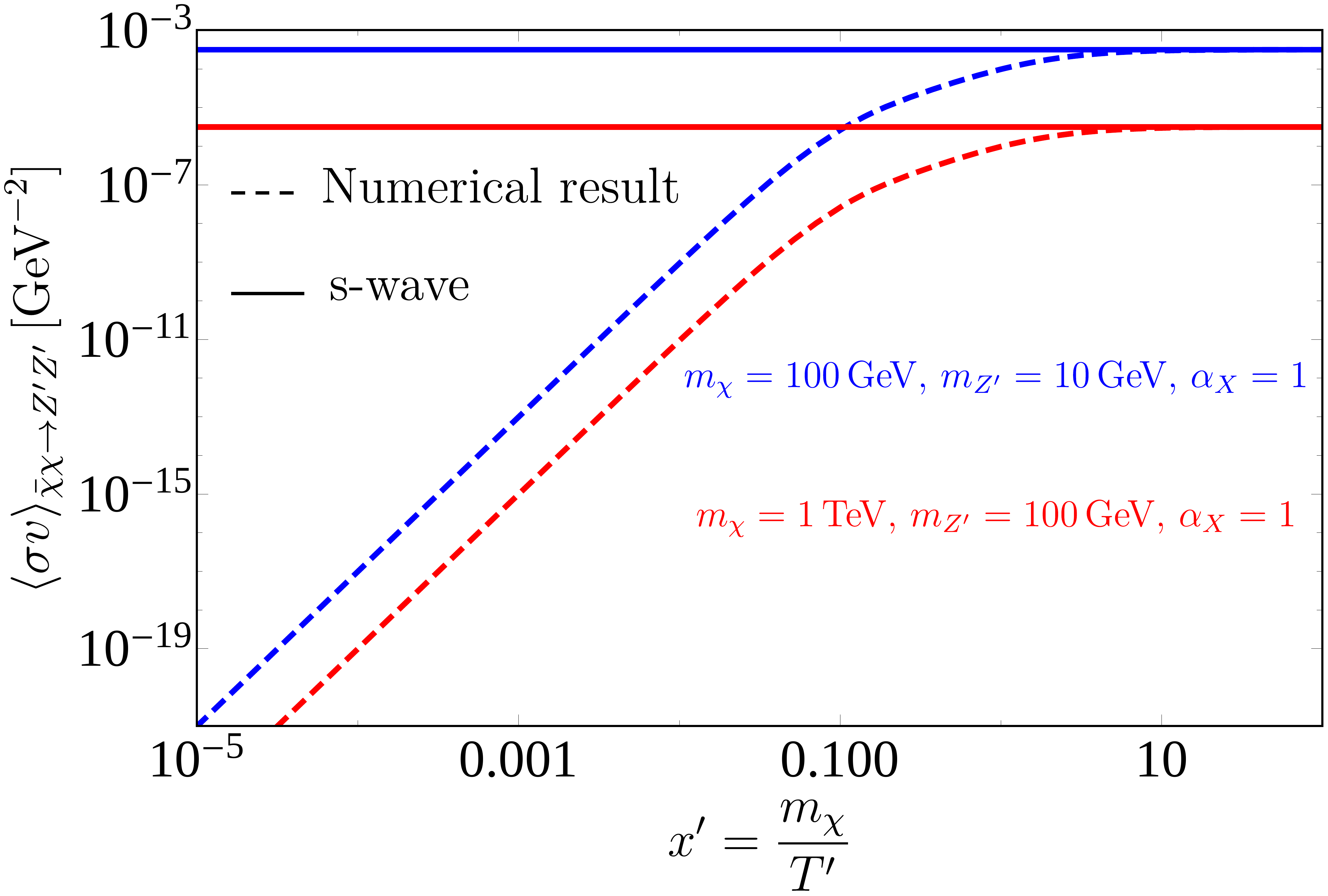}
\caption{Variation of $\langle \sigma v\rangle_{\bar{\x}{\x}\to\zp \zp}$
as a function of $\xp$ for two benchmark values of $m_\x$, $m_{\zp}$, and $\alpha_X$. 
The dashed and the solid lines denote the numerical result and s-wave
term of $\langle \sigma v\rangle_{\bar{\x}{\x}\to\zp \zp}$ respectively.
The blue lines are plotted for 
$m_\x = 100 \,{\rm GeV}$, $m_{\zp} = 10 \, {\rm GeV}$, and $\alpha_X = 1$
whereas for the red lines we consider 
$m_\x = 1 \, {\rm TeV}$, $m_{\zp} = 100 \,{\rm GeV} $, and $\alpha_X = 1$.
}
\label{cross-plot}
\end{figure}

In Fig.\,\ref{cross-plot}, we have plotted the variation of
numerically calculated $\langle \sigma v \rangle_{\bar{\x}{\x} \to \zp \zp}$
as a function of $\xp = m_\x/\tp$ for two benchmark values of $m_\x$, $m_{\zp}$, and $\alpha_X$
along with the s-wave term of $\langle \sigma v \rangle_{\bar{\x}{\x} \to \zp \zp}$.
From the plot it is clear that the s-wave term
coincides with the numerically evaluated thermal average
for $\xp \gtrsim 1$. Since the DM freezes out at $\xp \sim 10$,
we can safely use the s-wave annihilation cross
section, given in Eq. \ref{cross-swave}.

The second term on the right hand side of Eq. \ref{BE_n} is responsible for the production of 
DM from SM bath and $\langle \sigma v \rangle^{T}_{\bar{f}{f} \to \bar{\x} \x}$
is the thermally averaged cross section of $\bar{f}{f} \to \bar{\x}{\x}$ where
$f=\{\mu, \tau, \nu_\mu, \nu_\tau\}$ and it is
calculated at temperature $T$. Since the DM is not in thermal
contact with the SM bath therefore we have neglected the production of
SM particles from DM annihilation. The thermally averaged cross section
of $\bar{f}{f} \to \bar{\x}{\x}$ is defined as follows.
\bea
\label{Th-avg}
\langle \sigma v \rangle_{\bar{f}{f} \to \bar{\x}{\x}}
&=&
\dfrac{1}{n_{f_{\rm eq}}^2(T)}
\dfrac{g_f^2 T}{2 (2 \pi)^4}
\int_{{\rm Max}\left[4 m_\x^2, 4 m_f^2\right]}^\infty
\sigma_{\bar{f}{f}\to \bar{\x}{\x}} \sqrt{\hat{s}} \left(\hat{s} - 4 m_f^2\right)
K_1\left(\dfrac{\sqrt{\hat{s}}}{T}\right) \, d\hat{s}\,\,.
\eea 
Here $n_{f_{\rm eq}} (T)$ is the equilibrium number density of the
annihilating fermion $f$ at temperature
$T$, $g_f$ is the internal degrees of freedom of $f$, and
$\hat{s}$ is the Mandelstam variable. 
In Eq. \ref{Th-avg},
$K_1(\sqrt{\hat{s}}/T)$ is the modified Bessel function of second kind and order one.
The analytical form of the annihilation cross sections 
of $\bar{f}f \to \bar{\x} {\x}$
are given by
\bea
\sigma_{\bar{f}{f}\to \bar{\x}{\x}}
&=&
\dfrac{\alpha_X \eps^2}{3 \hat{s}}\sqrt{\dfrac{\hat{s} - 4 m_\x^2}{\hat{s} - 4 m_f^2}}
\dfrac{(\hat{s} + 2 m_f^2)(\hat{s} + 2 m_\x^2)}{(\hat{s} - m_{\zp}^2)^2} 
~~~~~ \text{for } f = \mu,\,\tau\,\,,\\
\sigma_{\bar{\nu}_i{\nu_i}\to \bar{\x}\x}
&=&
\dfrac{\alpha_X \eps^2}{6}
\sqrt{\dfrac{\hat{s} - 4 m_\x^2}{\hat{s}}}
\dfrac{\hat{s} + 2 m_\x^2}{(\hat{s} - m_{\zp}^2)^2}
~~~~ \text{for } \nu_i = \nu_\mu,\,\nu_\tau\,\,. 
\eea

In Eq. \ref{BE_n}, $n_{{\x_{\rm eq}}_{\rm tot}} (\tp)$ and 
$n_{{\x_{\rm eq}}_{\rm tot}} (T)$
are the equilibrium number densities of $\x + \bar{\x}$ 
calculated at temperatures $\tp$ and $T$
respectively. Let us note that the 1/2 (2) factor in the first (second) term on the right
hand side of this equation arises because of the fact that the DM
candidate is a Dirac fermion. 

To solve Eq. \ref{BE_n} we define the comoving number
density $Y_{\x_{\rm tot}}$ as $Y_{\x_{\rm tot}} = n_{\x_{\rm tot}}/s(T)$
where $ s(T) = \dfrac{2 \pi^2}{45} g_{*s} (T) T^3$ 
is the entropy density of the SM bath and $g_{*s}(T)$
is the relativistic degrees of freedom contributing
to the entropy density of the SM bath.
Now defining $x = m_\x/T$ and using the conservation
of entropy of the SM bath (approximately), we can arrive
at the following equation.
\bea
\dfrac{d Y_{\x_{\rm tot}}}{d x}
&=&
\dfrac{h_{\rm eff}(x)}{2}\dfrac{s(x)}{x H(x)}
\langle \sigma v\rangle ^{\tp}_{\bar{\x}\x\to \zp \zp}
\left(Y_{\x_{{\rm eq}_{\rm tot}}} (\tp,T)^2 - Y_{\x_{\rm tot}}^2\right)
\nn\\
&+&
\dfrac{2\,h_{\rm eff}(x) s(x)}{x H(x)}
\sum_f \langle \sigma v\rangle^T_{\bar{f}{f}\to \bar{\x}\x}
Y_{f_{{\rm eq}}}^2(T)
\,\,,
\label{BE_Y}
\eea
where $Y_{\x_{{\rm eq}_{\rm tot}}}(T,\tp) = \dfrac{n_{\x_{{\rm eq}_{\rm tot}}} (\tp)}{s(T)}$, 
$Y_{f_{\rm eq}} (T)= \dfrac{n_{f_{\rm eq}}(T)}{s(T)}$
and $h_{\rm eff} (x)= \left(1 - \dfrac{1}{3} \dfrac{d \ln g_{*s}(x)}{d \ln x}\right)$.

Now in the right hand side of Eq. \ref{BE_Y}, if the first term is dominant over the second one
i.e. if the freeze-in term is negligible then comoving number density of DM
is only governed by the annihilation cross section of $\bar{\x} \x \to \zp \zp$
and also the dark sector temperature $\tp$ which depends on $\eps$ and $T$. 
Thus in this case DM freezes out at $x^\prime_{\rm FO} \sim 10$ and 
this is known as the ``LIDM'' scenario. However if the 
freeze-in term is not negligible then in that case after decoupling of the DM, 
its comoving number density can be increased due to the DM production 
from SM bath and the production freezes at $x_{\rm FI} \sim 1$. In this
case the final relic density of DM does not depend on the first term of Eq.\,\ref{BE_Y}
and it is known as ``freeze-in". Thus the final relic density is independent  
of the $\bar{\x} \x \to \zp \zp$ annihilation cross section. 
In case of reannihilation, after the departure from the equilibrium number density, 
the DM produced from the SM bath due to the presence of the freeze-in term
can reannihilate into $\zp$ if the dark sector coupling is sufficiently large. 
Therefore in this case both the term on the right hand 
side of Eq. \ref{BE_Y} contribute to the final relic abundance.

Depending on the time ordering of $x_{\rm FO}$ 
(corresponding to $x^{\prime}_{\rm FO}\sim 10$) and $x_{\rm FI}$
the LIDM scenario can be classified into two parts:

i) If $x_{\rm FO} < x_{\rm FI}$ then freeze-in term is triggered after the 
DM freezes out and in this case the final relic abundance slightly depends
on the freeze-in term. This is known as ``early-LIDM",

ii) If $x_{\rm FO} > x_{\rm FI}$ then freeze-out occurs after
the production of DM stops. In that case since DM follows the equilibrium
number density during the production via freeze-in therefore
the produced DM can rapidly thermalise with the $\zp$ bath.
Thus the final relic abundance is independent of the freeze-in
term. This is known as ``Late LIDM".

In the next section we show our numerical results for
each of the above mentioned scenarios and also
the allowed parameter space for LIDM, freeze-in, and 
reannhilation.

\section{Numerical results for Dark Matter relic density}
\label{numerical_results}
In this section we discuss the numerical results
of our analysis. We have solved the Boltzmann equation
for the total number density of the DM candidate
by considering the fact that the dark sector evolves
non-adiabatically and its temperature evolution
is given by the semi-analytic result in Eq. \ref{xi-approx}.
Using the solution of the Boltzmann equation we can compute
the final relic abundance of $\x$ from the following relation.
\bea
\label{relic_den}
\Omega h^2 = 2.755\times 10^8 \left(\dfrac{m_\x} 
{1\,\rm GeV}\right) Y^0_{\x_{\rm tot}}\,\,,
\eea
where $Y^0_{\x_{\rm tot}}$ is the present value of the DM
comoving number density.
The initial conditions for solving the Boltzmann
equation are $x = x_{in}= 10^{-4}$ and 
$Y_{\x_{\rm tot}} =Y_{\x_{{\rm eq}_{\rm tot}}} (T_{in}, \tp_{in})$
where $T_{in} = m_\x/x_{in}$ and $\tp_{in} = \xi (T_{in}) T_{in}$.
In choosing the initial condition we have assumed the DM
follows the equilibrium distribution at $x = x_{in}$ 
and the validity of this assumption will be 
discussed in Appendix \ref{App:D}. 
Let us note that in our numerical analysis
we choose $r = m_{\zp}/m_\x = 0.1$ for the calculation of 
the DM relic density and parameter spaces allowed from
the relic density constraint remain unaltered for other
choices of $r$. The only required condition is
at the time of DM freeze-out
$\zp$ should behave as radiation and only in that case we can
write its energy density proportional to ${\tp}^4$. Therefore
at the time of DM freeze-out $\tp_{f} \gtrsim m_{\zp}$ which
implies $r^{-1} \gtrsim 10$ (considering $\xp_f \sim 10$).
We would also like to mention that in our numerical analysis 
we have used Eq. \ref{xi-approx} in the expression of $Y_{\x_{{\rm eq}_{\rm tot}}}$.
Since this relation is valid for $T\gtrsim10 m_{\zp}$, 
we can write this relation as $r^{-1}\gtrsim 10 \,x$ 
where we have used $T = m_\x/x$. Since DM
freezes-out as $x<1$ therefore this condition is satisfied
for $r \leq 10^{-1}$. Thus we can safely use Eq. \ref{xi-approx}
for the expression of $\tp$.

\begin{figure}
\centering
\includegraphics[height = 8cm, width = 8cm]{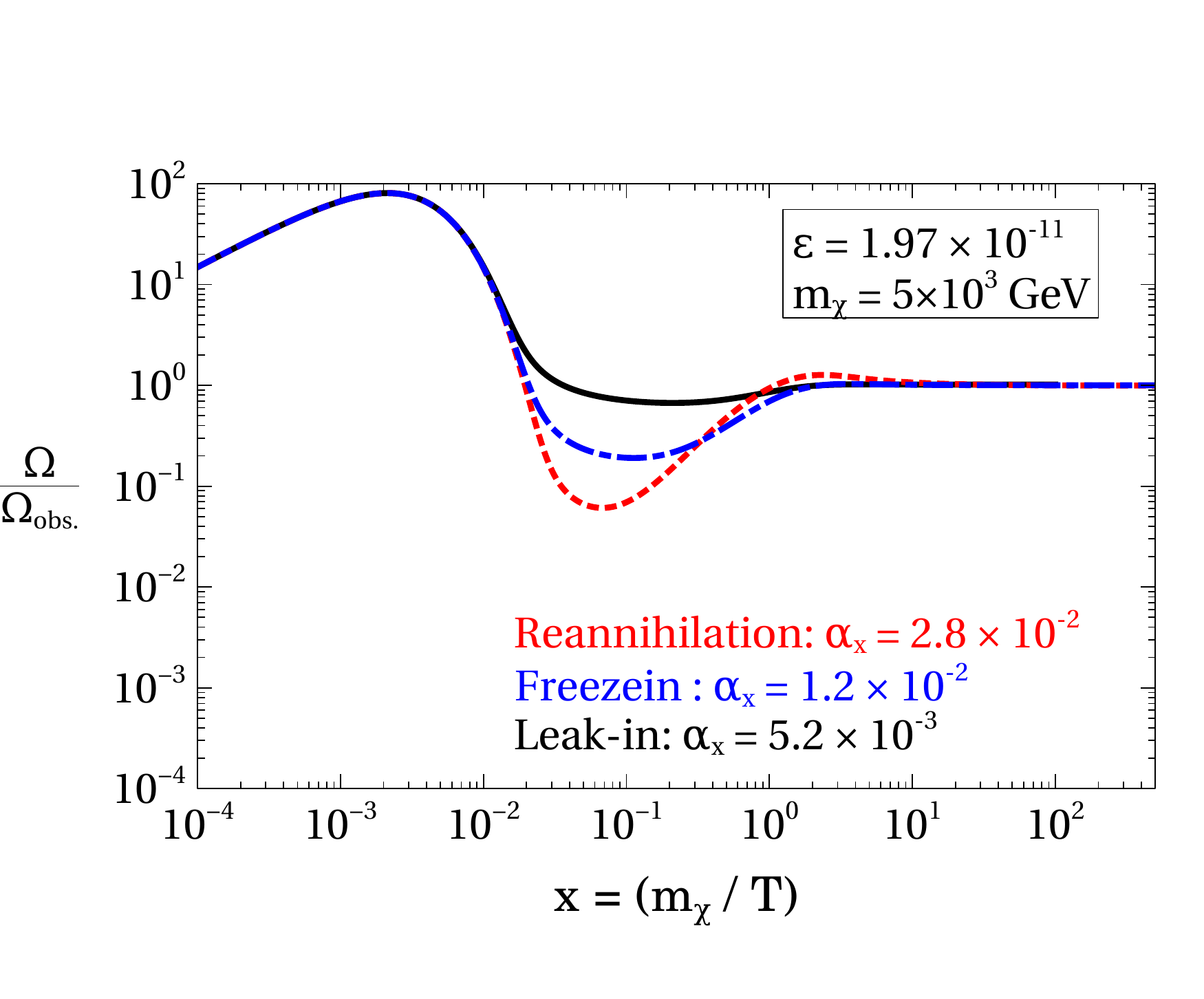}
\includegraphics[height = 8cm, width = 8cm]{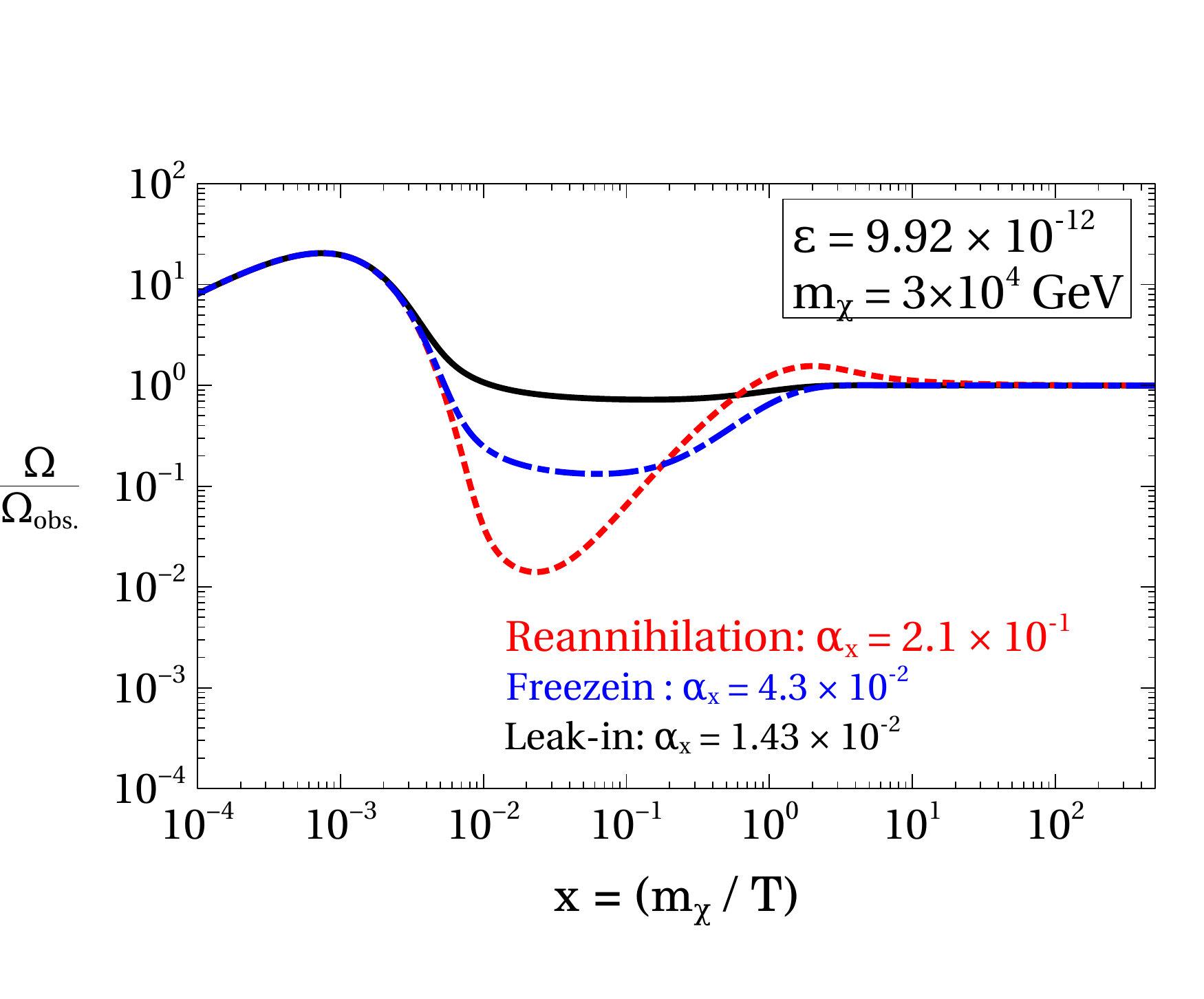}
\caption{\textbf{\textit{Left panel:}} The variation of relative abundance of DM
as a function of $x$ for $m_\x = 5\times 10^3 \,{\rm GeV}$, $\eps = 1.97 \times 10^{-11}$ and for
three difference choice of $\alpha_X$. The three different mechanisms i.e.
early leak in ($\alpha_X = 5.2 \times 10^{-3}$), freeze-in ($\alpha_X = 1.2 \times 10^{-2}$),
and reannihilation ($\alpha_X = 2.8 \times 10^{-2}$) are depicted by solid black, blue dashed-dot, and
red dotted lines respectively.
{\textbf{\textit{Right panel:}}} The variation of relative abundance of DM
as a function of $x$ for $m_\x = 3\times 10^4 \,{\rm GeV}$, $\eps = 9.92 \times 10^{-12}$ and for
three difference choice of $\alpha_X$. Here,
early leak in ($\alpha_X = 1.43 \times 10^{-2}$), freeze-in ($\alpha_X = 4.3 \times 10^{-2}$),
and reannihilation ($\alpha_X = 2.1 \times 10^{-1}$) are depicted by solid black, blue dashed-dot, and
red dotted lines respectively. In both the panel we have considered $m_{\zp} = 0.1 m_{\x}$.
}
\label{LI-FI-RE}
\end{figure}

In Fig.\,\ref{LI-FI-RE} we show the evolution of the relative
abundance of the DM (defined as the ratio between the 
DM abundance evaluated from Eq. \ref{relic_den} to the observed
value of the DM relic density ($\Omega h^2_{\rm Obs.} \simeq 0.12$)) 
as a function $x$ for two benchmark 
values of the model parameters. In each of the plots
we fix the value of $m_\x$ and $\eps$ and vary $\alpha_X$
to get the final relic abundance. Let us consider the
left panel of Fig.\,\ref{LI-FI-RE}. In this figure, for $\alpha_X = 5.2 \times 10^{-3}$, 
the DM decouples at $x \sim 10^{-2}$ from dark radiation 
bath and its abundance increases slightly due to the presence of
the freeze-in term in the Boltzmann equation. The increase in the
DM abundance stops at $x \sim 2$ and after that it remain constant.
This is known as the `early-LIDM' as discussed earlier and it
is denoted by the solid black line. Now if we increase
$\alpha_X$, the results of the increase are two fold.
Firstly with the increase in $\alpha_X$ the DM interacts with
the radiation bath more strongly and there will be a delay 
in the decoupling of the DM. Secondly for large values of $\alpha_X$
the freeze-in term plays a crucial role since it is proportional to 
$\alpha_X$ and depending on the values of $\alpha_X$ the final DM
abundance is either set by freeze-in mechanism or reannihilation.
From the figure one can see that for $\alpha_X = 1.2 \times 10^{-2}$
the final abundance of the DM is set only by the production
of the DM from SM bath and the production freezes at $x \sim 2$.
Thus in this case the first term in the right hand side of Eq. \ref {BE_Y}
does not play any significant role. This is known as Freeze-in mechanism
which is depicted by blue dashed-dot line. In contrast to that
for $\alpha_X = 2.8 \times 10^{-2}$ 
the production of DM from the SM bath is sufficient 
to rethermalise the DM with $\zp$. Therefore both the
term in the right hand side of Eq. \ref{BE_Y} determine
the final relic abundance and reannihilation occurs.
The red dotted line indicates the reannihilation mechanism
of DM. In the right panel of Fig.\,\ref{LI-FI-RE} we show all of the
three mechanisms discussed earlier with the same color codes 
but with different choice of the parameters. Let us note that
for both the plots we choose $m_{\zp} = 0.1 m_{\x}$ and the results
shown in Fig.\,\ref{LI-FI-RE} are independent of this choice
as long as $m_{\zp} \le 0.1 m_{\x}$.

\begin{figure}
\centering
\includegraphics[height=7cm,width=7cm]{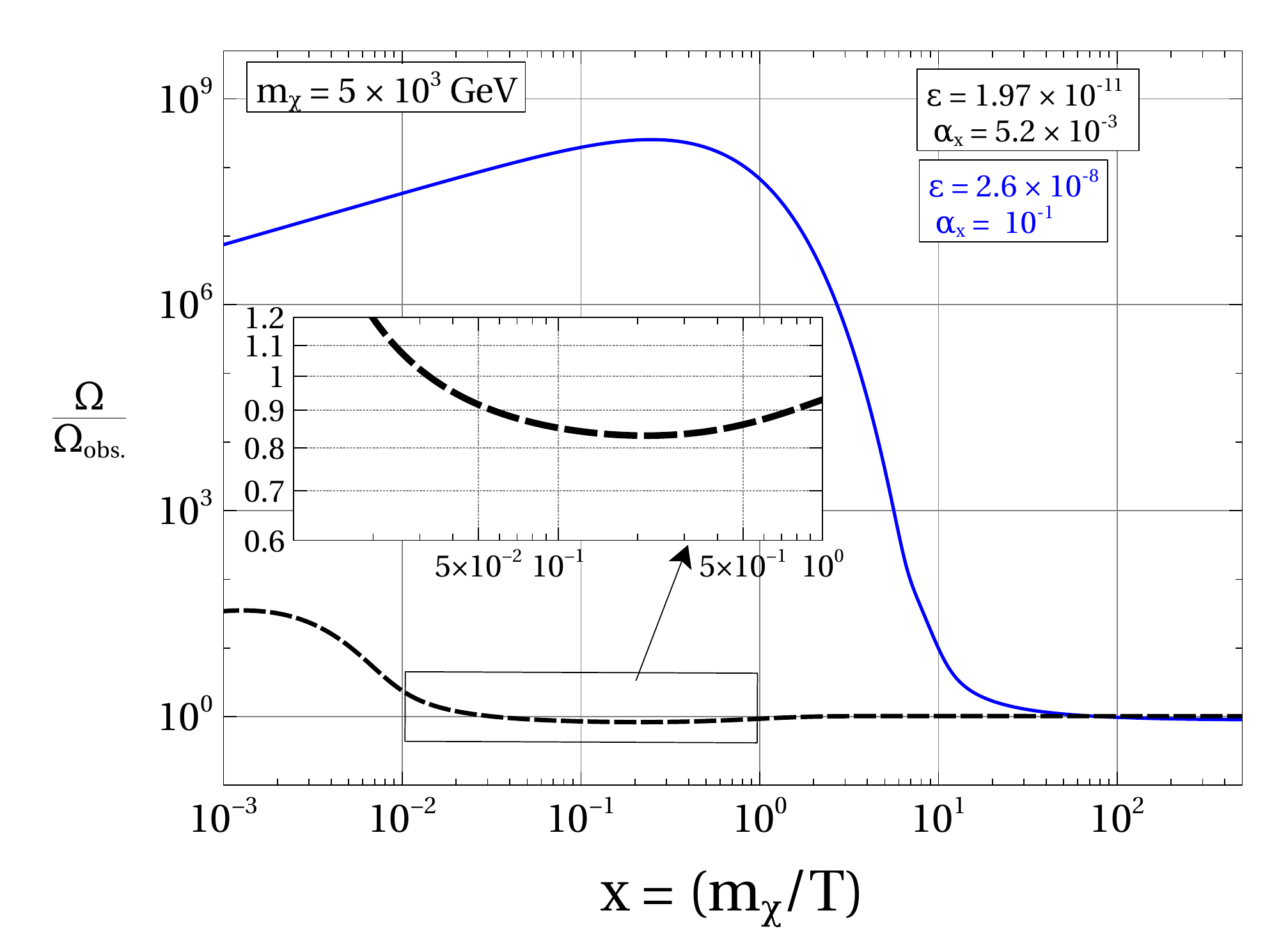}
\includegraphics[height=7cm,width=7cm]{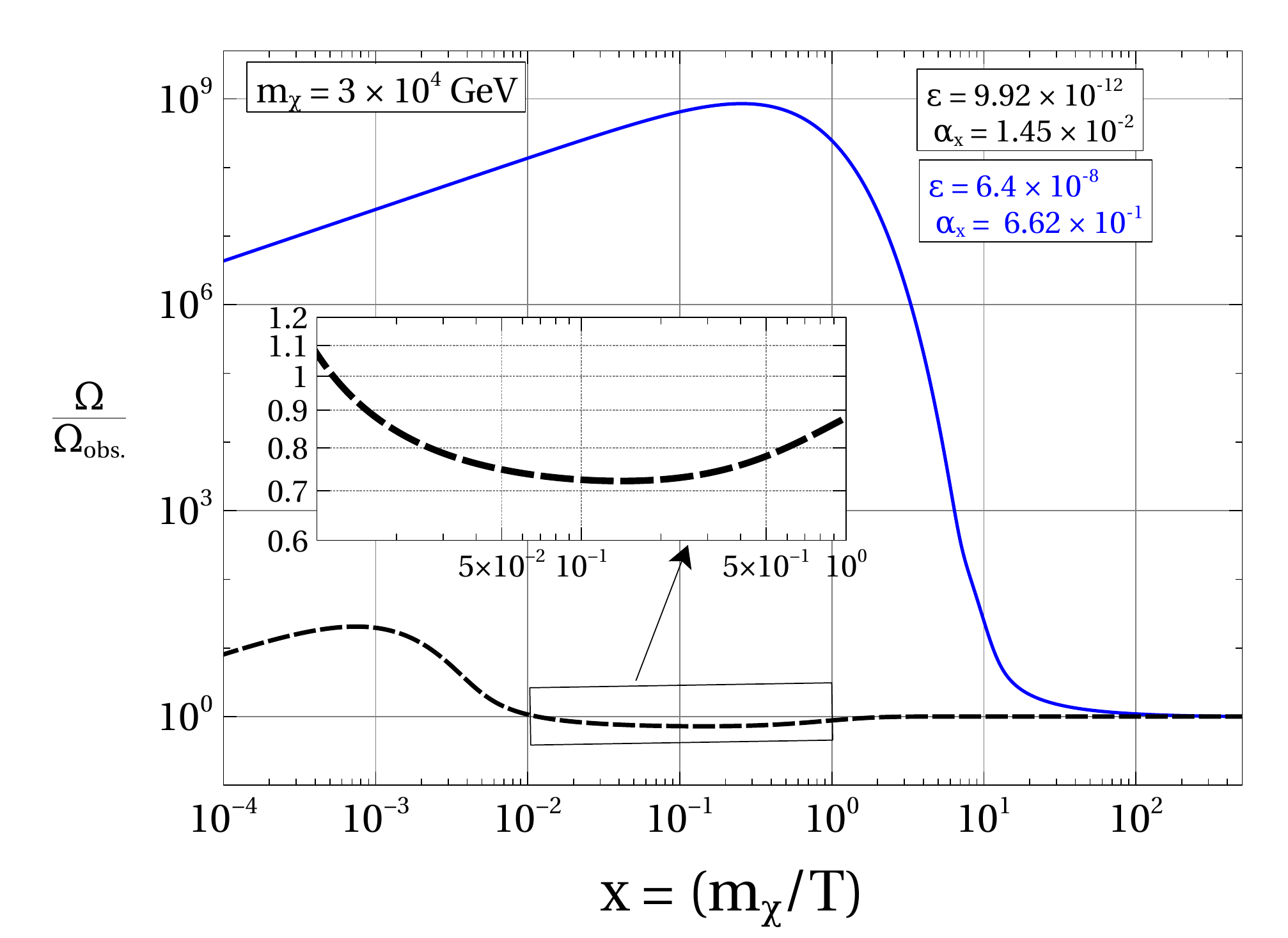}
\caption{\textbf{\textit{Left panel:}}
Evolution of the relative DM abundance as a function of $x$ 
in case of late and early LIDM for $m_{\x} = 5 \times 10^{3}\,{\rm GeV}$. 
The late-LIDM mechanism
is shown by the blue solid line and the values of $\eps$ and $\alpha_X$
are $2.6\times 10^{-8}$ and $10^{-1}$ respectively. The black dashed
line depicts the early-LIDM scenario for $\eps = 1.97 \times 10^{-11}$
and $\alpha_X = 5.2 \times 10^{-3}$.
\textbf{\textit{Right panel:}} DM relative abundance as a function of 
$x$ for $m_\x = 3 \times 10^{4}\,{\rm GeV}$. The color codes are same
as left panel but the corresponding values of $\eps$ and $\alpha_X$
are different 
(for late-LIDM $\eps = 6.4 \times 10^{-8}$, 
$\alpha_X = 6.8 \times 10^{-1}$ (blue solid line),
and for early-LIDM $\eps = 9.92 \times 10^{-12}$, 
$\alpha_X = 1.45 \times 10^{-2}$ (black dashed line)). 
In both the panel the value of $m_{\zp} = 0.1 m_{\x}$.
}
\label{EL-LL}
\end{figure}

As we discussed earlier, depending on the time ordering
of $x_{\rm FO}$ and $x_{\rm FI}$ LIDM scenario can be classified
into early-LIDM and late-LIDM. In Fig.\,\ref{EL-LL} we show
the evolution of the relative DM abundance as a function of $x$
for both the LIDM scenarios. Since the equilibrium comoving number density
of the DM increases with the increase in $\eps$ therefore for 
large values of $\eps$ we need large annihilation cross section
to get the correct relic abundance and thus the DM
is in thermal contact with the dark radiation bath 
for longer time. In this case the production
of DM from SM bath does not play any significant role
if the DM is in thermal contact with dark radiation bath. 
This is because the DM produced from SM bath will rapidly 
annihilate into $\zp$. Thus the comoving number density
of DM remains unaltered despite the fact that the
freeze-in term of the Boltzmann equation is not negligible.
This scenario is known as `late-LIDM' and blue solid line
of each panel of this figure indicates the DM coving number density
for late-LIDM scenario.
 
However the situation is different for small values of $\eps$ and $\alpha_X$. In this
case the DM decouples much earlier compared to late-LIDM
scenario and the production of DM from the SM bath occurs
after the decoupling of DM. Therefore the final abundance 
of the DM increases slightly at $x\sim 2$. This phenomenon 
is called as `early-LIDM'. In both the panel the black 
dashed lines depict the early-LIDM scenario.

\begin{figure}
\centering
\includegraphics[height = 10cm, width = 12cm]{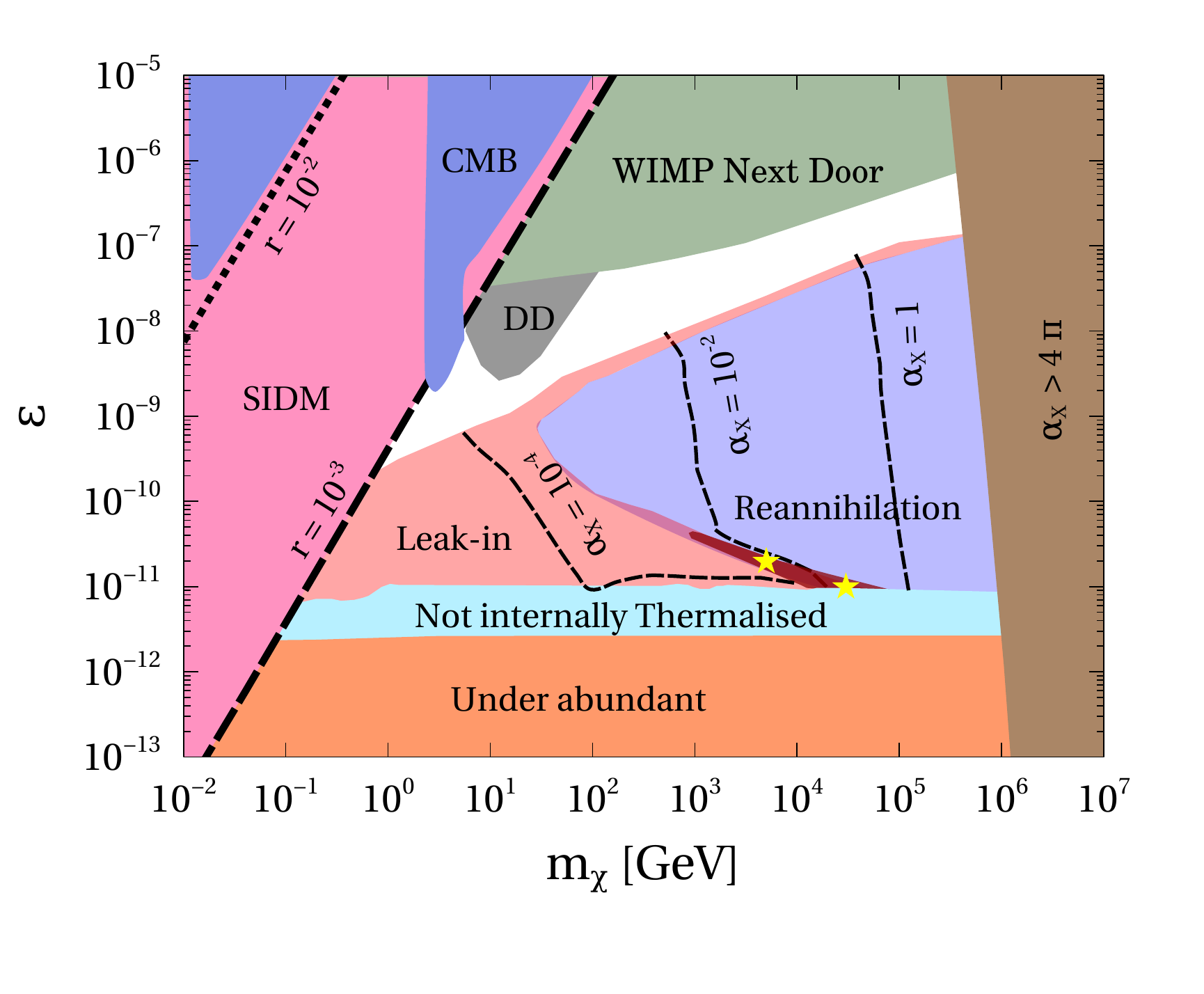}
\caption{Allowed parameter space in $m_\x -\eps$ plane
where the contours of $\alpha_X = 10^{-4}, 10^{-2}, 1$ are shown with
the dashed lines. The points mentioned in Fig.\,\ref{LI-FI-RE} are shown with
yellow marked points at which leak-in, freeze-in, and reannihilation occur.
Parameter space for LIDM is shown with light red color whereas
purple region represents the parameter space for reannihilation.
The small red patch corresponds to the region where all of 
the three non-equilibrium stages
namely leak-in, freeze-in, and reannihilation are possible.
The green region is the parameter space for WIMP next door scenario and the region
in which $\alpha_X > 4\pi$ is shown by the brown color. Cyan region
shows the parameter space for not internally thermalised dark sector
(considering $m_{\zp}/m_{\x} = 10^{-6}$)
whereas in the orange region the number density of the DM
is not sufficient to produce correct relic abundance.
The constraint from self interacting dark matter (SIDM)
is shown by the pink color for two values of
$r$ ($r = 10^{-2}$ (dotted line), $r = 10^{-3}$ (dashed line)).
They grey and blue region show the direct detection
and CMB constraint for $r = 10^{-3}$ and $r = 10^{-1}$ respectively.
\label{sum-eps-mdm}
}
\end{figure}

Finally in Fig.\,\ref{sum-eps-mdm} we show the allowed parameter space
in $m_\x -\eps$ plane from relic density along with
other experimental constraints. In this plot
we show the contours of $\alpha_X = 1, 10^{-2}, 10^{-4}$ with
black dashed lines. Because the dark sector temperature
increases with the increase in $\eps$, we need
higher values of annihilation cross section to 
satisfy the correct relic abundance. This implies that
for constant $\alpha_X$, $m_\x$ should decrease as we
increase $\eps$. One can clearly see this pattern
for each contour of constant $\alpha_X$.
We can see for a fixed value of $\eps$, $m_\x$ increases
with the increase in $\alpha_X$. This behaviour can be
understood as follows. 
For a fixed value of $\eps$, annihilation
cross section increases due to the increase in $\alpha_X$
and this enhancement in the cross section should be counterbalanced
by the increase in $m_\x$ to satisfy the correct relic density. 
However for lower values of $\eps$
i.e. $\eps \lesssim 10^{-11}$ all the contours of constant $\alpha_X$
coincide and in that case for fixed value of $m_\x$ and $\eps$ one can obtain
many relic density satisfied points for different values of $\alpha_X$.
Thus in this region, for a fixed value of $m_\x$ and $\eps$,
the mechanisms of getting correct relic abundance
are different and they depend on the values of $\alpha_X$.
The purple and light red region in this figure
denote the relic density satisfied region
via reannihilation and leak-in mechanisms respectively.
The red patch in the parameter space denotes the region
at which leak-in, freeze-in, and reannihilation occurs.
In Fig.\,\ref{LI-FI-RE} we showed the leak-in, freeze-in, and reannihilation
for two fixed values of $m_\x$ and $\eps$ and these two points
are shown with yellow marked points in Fig.\,\ref{sum-eps-mdm}.

As mentioned earlier the dark sector is not in thermal equilibrium
with the SM bath therefore we have identified the parameter space
for this scenario according to the discussion in section \ref{wimp-next-door}. 
The green region of Fig.\,\ref{sum-eps-mdm} indicates the allowed region
for the WIMP next door scenario. The parametric dependence
of the $\Gamma/H$ ratio is given by $\eps^2/m_\x$. Thus in this region
$\eps$ increases with the increase in $m_\x$ and this dependence 
can be seen clearly from the figure. 

The equilibrium number density of the DM depends on
the dark sector temperature $\tp$ which is 
proportional to $\eps^{1/2}$. Thus equilibrium number density 
decreases with the decrease in $\eps$.
This means there will be a minimum value of $\eps$ below
which it is not possible to get the correct relic abundance.
The orange region of Fig.\,\ref{sum-eps-mdm} shows the minimum values
of $\eps$ below which we cannot obtain correct relic abundance.
This region has been obtained from the condition 
$Y_{\x_{{\rm eq}_{\rm tot}}}|_{\rm max} (\tp_0)\leq \left(0.12/2.755 \times 10^8\right) 
(1 \,{\rm GeV}/m_\x)$ where $Y_{\x_{{\rm eq}_{\rm tot}}}|_{\rm max} (\tp_0)$ is the maximum
value of the comoving number density of DM and $\tp_0 = 2 m_{\x}/5$ is the value
of the dark sector temperature at which maximum occurs \cite{Evans:2019vxr}.

The brown region of the figure represents that the required
values of $\alpha_X$ for correct relic density 
belongs to the non-perturbative regime of $\alpha_X$
i.e. in this region of parameter space $\alpha_X > 4 \pi$.

We have also shown the constraints from internal thermalisation
of the dark sector by cyan region and the relevant discussion is given
in Appendix \ref{App:D}. There are other 
constraints such as self interactions of  dark matter (pink region),
direct detection (DD) (grey region), CMB (blue region).
In deriving these constraints we express $\alpha_X$ in terms of $\eps$ and $m_\x$
(this can be done by using Eq. \ref{xi-approx},
Eq. \ref{relic_approx} and Eq. \ref{cross-swave}) to identify the parameter space only for LIDM.
In section \ref{DM-indi} we will discuss these constraints in detail and
we will also discuss the allowed parameter space in $m_\x -\alpha_X$ 
plane by considering $\alpha_X$ and $\eps$ as independent parameters.

\section {Detection prospect of the Dark Matter}
\label{DM-indi}
The parameter space of our model can be constrained from
various astrophysical and laboratory experiments. In this
section we will discuss the constraints on the model
parameters such as $\eps$, $m_\x$, $m_{\zp}$, and $\alpha_X$.
\subsection{Direct Detection}
\label{DD}
In our model DM interacts with SM particles (namely, second and 
third generation leptons) through the kinetic
mixing of $\zp$ and $Z_{\mt}$. Although there is no
tree level interaction of DM with
quarks but still interaction can happen  through
the radiatively generated $Z-\zp$ and $\gamma-\zp$ 
kinetic mixing\footnote{
Here we have not considered the DM-nucleon scattering
via radiatively induced $Z-\zp$ kinetic mixing $\Pi(q^2)_{Z\zp}$ because 
the coupling of SM quarks with $\zp$ via $Z-\zp$ mixing is proportional 
to $ \Pi(q^2)_{Z\zp} \dfrac{m_{\zp}^2}{m_Z^2}$ where
$m_Z$ is the mass of the SM $Z$ boson \cite{Bauer:2018onh}
and in our parameter space of interest ${m_{\zp}^2}\ll{m_Z^2}$.}. 
Therefore we have calculated 
spin independent DM-nucleon scattering cross section to study
the direct detection constraint and the scattering
cross section in the low momentum transfer limit
is given by
\bea
\label{DD}
\sigma^{\rm SI}_{\x n} &=&
\frac{4 \alpha_X \mu_{\x n}^2}{m_{\zp}^4}
|\Pi(0)|^2
\left[\frac{Z}{A}f_p +\left(1-\frac{Z}{A}\right)f_n\right]^2\,\,,
\eea
where $\mu_{\x n} = \frac{m_n m_\x}{m_n + m_\x}$ is the reduced mass of the
DM-nucleon system and $m_n$ is the mass of a nucleon.
$A(Z)$ is the mass (atomic) number of target nucleus. $f_p$ and
$f_n$ are defined as follows \cite{Berlin:2014tja}: 
$f_p = 2 Q_u + Q_d$ and
$f_n = Q_u +2 Q_d$, where $Q_u \left(Q_d\right)$ is 
electromagnetic charge of up (down) quark. The radiatively
generated kinetic mixing between $\zp$ and $\gamma$
at one loop level is given by \cite{Araki:2017wyg,Banerjee:2018mnw}
\bea
\label{mixing-loop}
i \Pi (q^2) = - \frac{i e \epsilon}{2 \pi^2}
\int_0^1 x (1-x) \ln\left(\dfrac{m_\mu^2 -q^2 x(1-x)}
{m_\tau^2 - q^2 x(1-x)} \right) dx\,\,\,,
\eea
where $e = \sqrt{4 \pi \alpha_{\rm em}}$ is the charge
of the electron and $\alpha_{\rm em}$ is the fine-structure
constant. In the limit $q^2 \to 0$ Eq. \ref{mixing-loop} takes
the following form.
\bea
\label{mixing-loop-approx}
i \Pi (0) = - \frac{i e \epsilon}{12 \pi^2} 
\ln\left(\dfrac{m_\mu^2}{m_\tau^2}\right)\,\,.
\eea

Moreover we have also considered the DM-electron scattering
for low mass DM and the corresponding scattering cross section
in the low momentum transfer limit is given by
\bea
\label{DD_e}
\sigma_e &=&
\dfrac{16 \pi \alpha_X \alpha_{\rm em}\, \mu_{\x e}^2}{m_{\zp}^4}
|\Pi(0)|^2\,\,,
\eea
where $\mu_{\x e} = \frac{m_\x m_e}{m_\x + m_e}$ is the reduced
mass of the DM-electron system.

Now we have compared the spin independent direct detection 
cross sections given in Eq. \ref{DD} and Eq. \ref{DD_e} with
the current bounds from the direct detection experiments 
XENON1T \cite{XENON:2018voc}, 
CRESST-III \cite{CRESST:2019jnq}, 
and XENON10T \cite{XENON10:2011prx,Essig:2017kqs}. The excluded
region of the parameter space in $m_\x - \alpha_X$ plane 
from direct detection constraint for $\eps = 10^{-9}$ and $r= m_{\zp}/m_\x= 10^{-3}$
is shown in Fig.\,\ref{alpha-mx} by yellow color and this bounds
will be more stringent for smaller values of $r$.

\subsection{CMB constraint}
\label{CMB}
Production of SM charged particles from the annihilation of DM 
can alter the ionisation history of Hydrogen
at the time of CMB. The amount of energy injected
per annihilation within the volume $\Delta V$ and time $\Delta t$ is given by
\bea
\Delta E = 2 m_\x \left(\dfrac{n_\x}{2}\right)^2_{\rm CMB} 
\langle \sigma v\rangle_{\bar{\x}{\x}\to \rm SM}
\Delta V \Delta t\,\,\,,
\eea
where $n_{\x}$ is the number density of DM at the time of CMB and 
$\langle \sigma v\rangle_{\bar{\x}{\x}\to \rm SM}$
is the annihilation cross section of DM into SM particles.
Now using the conservation of total number of DM we can write the expression
of the amount of deposited energy as follows.
\bea
\dfrac{d E}{dV dt} = (1+z)^6 \rho_c^2 \, \Omega h ^2 \,\mathcal{P}_{\rm ann}\,\,\,,
\eea
where\footnote{For self-conjugate DM the $\frac{1}{2}$ factor 
will be absent from the expression of $\mathcal{P}_{\rm ann}$.}
$\mathcal{P}_{\rm ann} = \dfrac{f_{\rm eff} (m_\x)}{2}\dfrac{\langle \sigma 
v\rangle_{\bar{\x}{\x}\to SM}}{m_{\x}}$, $z$ is the redshift parameter, and
$f_{\rm eff}$ is the redshift independent efficiency factor \cite{Slatyer:2015jla}.
For the $s$-wave DM annihilation the upper limit of $\mathcal{P}_{\rm ann}$ is
$4.1 \times 10^{-28}\rm \, cm^3 \, s^{-1} \, GeV^{-1}$ \cite{Planck:2015fie}. 
Therefore using the expression of $\mathcal{P}_{\rm ann}$ we can write
\bea
\label{CMB}
\sum_{\rm SM} f^{\rm SM}_{\rm eff} (m_\x) 
\dfrac{\langle \sigma v\rangle_{\bar{\x}\x\to \zp \zp} {\rm Br} (\zp \to \rm SM)}{m_\x}
< 8.2 \times 10^{-28}\rm \, cm^3 \, s^{-1} \, GeV^{-1}\,\,.
\eea
In deriving the CMB bound we have used Eq. \ref{CMB} and the corresponding
excluded region for $r = 10^{-1}$ is shown in Fig.\,\ref{alpha-mx} with red color.
The CMB bound for $r = 10^{-1}$ is maximum and it relaxes with the decrease in $r$
because of the reduction of the branching ratios of $\zp$ into SM particles.
\subsection{Dark matter self interaction}
\label{SIDM}
From the bullet cluster observations, the upper limit
of the self-interaction cross section of DM
is bounded as $\sigma_T/m_\x < 1.25\, {\rm  cm}^2 \, {\rm g}^{-1}$
\cite{Randall:2008ppe}
where $\sigma_T$ is the momentum transfer cross section
of DM-DM scattering process. To derive the allowed parameter space
from the bullet cluster observation, we have calculated
the momentum transfer cross sections of 
$\bar{\x} \x \ra \bar{\x} \x$, $\x \x \ra \x \x$, and
$\bar{\x} \bar{\x} \ra \bar{\x} \bar{\x}$ and defined
an effective cross section as \cite{Choi:2016tkj}
\bea
\label{SIDM}
\sigma_T = \dfrac{1}{4}\left(
\sigma^T_{\bar{\x} \x \ra \bar{\x} \x} +
\sigma^T_{\bar{\x} \bar{\x} \ra \bar{\x} \bar{\x}}+
\sigma^T_{\x \x \ra \x \x}
\right)\,\,,
\eea
where $1/4$ factor is due to the fact that
there is no asymmetry between particle and anti-particle,
therefore both of them contribute equally to the
relic density.

\begin{figure}
\centering
\includegraphics[height = 9cm, width = 11cm]{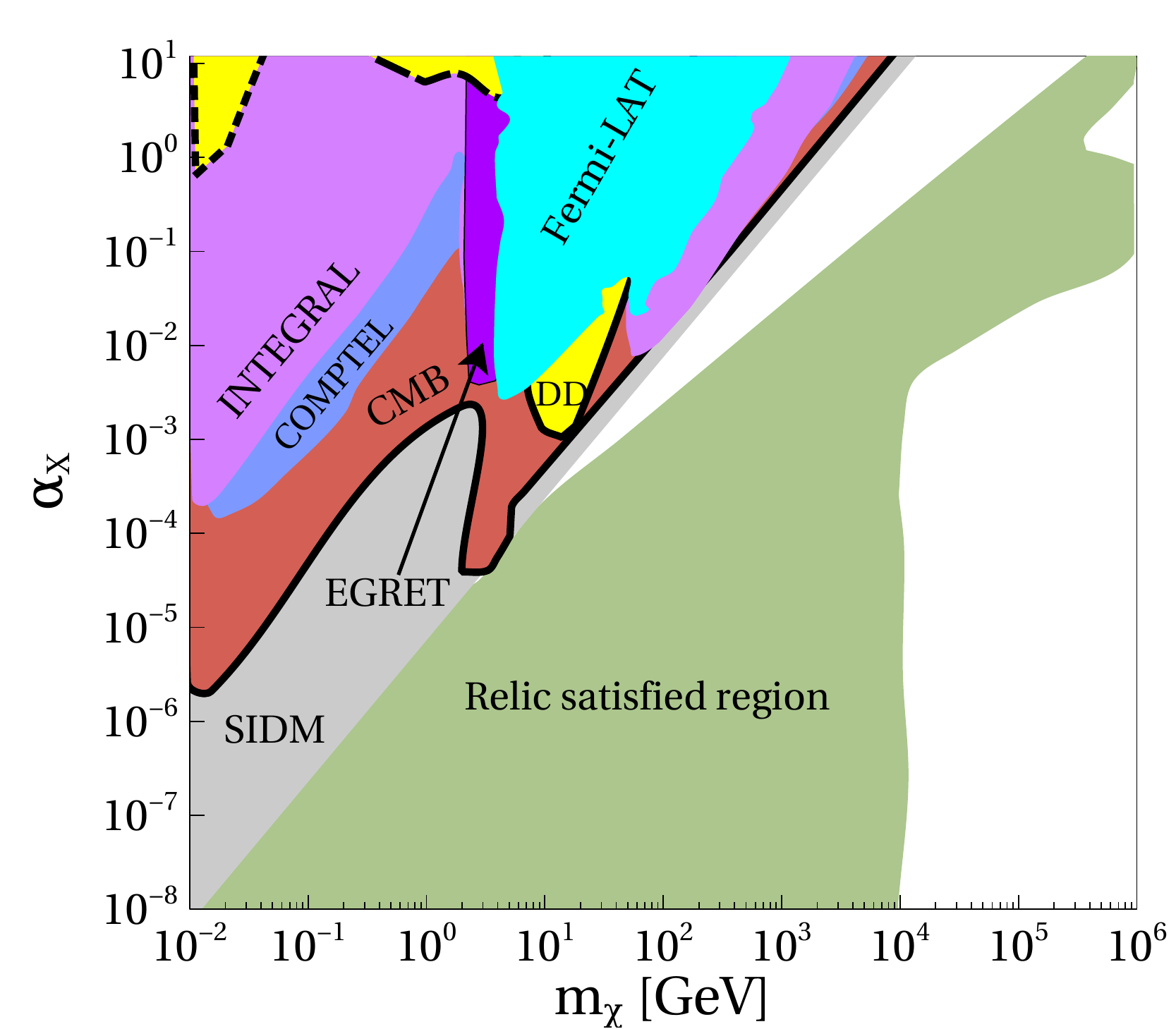}
\caption{Parameter space in $m_\x - \alpha_X$ plane. The green region is the
relic density satisfied region whereas the grey region is excluded
from the bullet cluster observation for $r = 10^{-3}$.
The red region is excluded from the observations of CMB for $r = 10^{-1}$.
The yellow region outlined with solid, dashed, and dotted line represent the direct detection
constraint from XENON1T, CRESST-III, and XENON10T for $\eps =10^{-9}$, and $r = 10^{-3}$.
The bounds arising from the measurement of diffuse $\gamma$-ray background flux 
(discussed below) by EGRET, INTEGRAL, COMPTEL, and Fermi-LAT are shown by
violet, purple, blue, and cyan regions respectively for $r = 10^{-1}$.
}
\label{alpha-mx}
\end{figure}

The bounds from self-interacting DM has been shown in Fig.\,\ref{alpha-mx}
by the grey region for $r = 10^{-3}$ and this bound will be stronger 
as we decrease $r$.

\subsection{$\gamma$-ray signal from DM annihilation}
\label{gamma_ray}
In our framework we will study the prospect of detecting $\gamma$ ray signal
from DM annihilation via one step cascade process\footnote{We 
have not considered the $s$ channel processes $\bar{\x}\x\to {\rm SM \, SM} $
since these processes are suppressed by $\eps^2$ at the cross section level. 
Note that throughout our analysis we have considered
$m_\x > m_{\zp}$ therefore resonance condition $m_{\zp} = 2 m_{\x}$
is not satisfied}. In this process, DM annihilates into a pair of $\zp$
and one of this $\zp$ can decay into a pair of SM charged particles.
The differential photon flux originating from this type of one step cascade
process is given by
\bea
\label{diff_flux}
\dfrac{d \Phi}{d E_\gamma \Delta \Omega}
&=&
\dfrac{\langle \sigma v \rangle _{\bar{\x}\x\to \zp \zp}}{16\pi m_\x^2}
\rho_\odot^2 R_\odot \bar{J}_{\rm ann}
\sum_{f} {\rm Br} (\zp \to \bar{f}f) \dfrac{d N^f_\gamma}{d E_\gamma}\,\,,
\eea
where $\dfrac{d N^f_\gamma}{d E_\gamma}$ is the photon spectrum
for the DM annihilation into $\bar{f}f$ pair where $f$ is a SM charged fermion
and the spectrum is calculated in the centre of mass (CoM) frame 
of the DM annihilation. $\langle \sigma v \rangle _{\bar{\x}\x\to \zp \zp}$
is the annihilation cross section of DM into a pair of $\zp$ and 
the s-wave term of the annihilation cross section is given in
Eq. \ref{cross-swave}. In Eq. \ref {diff_flux}, ${\rm Br} (\zp \to \bar{f}f )$
is the branching ratio of $\zp$ into a pair of SM fermions $\bar{f}f$ 
where $f = {e^-, \mu^-,\tau^-}$. Finally $\bar{J}_{\rm ann}$ 
is the average $J$ factor for DM annihilation
which is defined in the galactic co-ordinate 
system $(b,l)$ in the following way.
\bea
\label{J-fact}
\bar{J}_{\rm ann} = \dfrac{1}{\Delta \Omega}\int_{\Delta \Omega} \int_0^{\ell_{max}}
\left( \dfrac{\rho \left(\sqrt{\ell^2 + R_\odot^2 - 2 \ell R_\odot \cos b \cos l }\right)}
{\rho_\odot}\right)^2 \dfrac{d \ell}{R_\odot} d \Omega\,\,\,.
\eea 
Here $\rho_\odot = 0.3 \, {\rm GeV cm^{-3}}$ is the density
of DM at the solar location whereas $R_\odot = 8.3 \, {\rm kpc}$
is the distance between the galactic centre (GC) and the solar
location. $\rho (r)$ is the DM density profile and it is taken to 
be Navarro-Frank-White (NFW) density profile \cite{Navarro:1995iw} 
throughout our analysis.
$\ell$ is the line of sight ($l.o.s$) distance and the upper limit
of the $l.o.s.$ integration is given by
\bea
\label{lmax}
\ell_{\rm max} = 
\sqrt{R_{\rm MW}^2 - R_\odot^2 + R_\odot^2 \cos b\cos l} + R_\odot \cos b\cos l\,\,
\eea
where $R_{\rm MW} = 40 \, {\rm kpc}$ is the radius of the Milky Way (MW) galaxy.

Now in our scenario the $\gamma$ ray flux is composed of two components and
the components are i) prompt gamma ray from DM annihilation,
ii) secondary emission via inverse Compton scattering (ICS).

\begin{itemize}
\item \textbf {Prompt $\gamma$ ray: } DM annihilation into a pair of
$\zp$ and the subsequent decay of $\zp$ into charged SM fermions
can produce $\gamma$ via electroweak bremsstrahlung processes. This is
known as prompt gamma rays. For the spectrum of the prompt gamma rays, 
we have used the publicly available code 
\texttt{ PPPC4DMID} \cite {Cirelli:2010xx}
for $5  \,{\rm GeV} \le m_{\zp}/2\le 100\, {\rm  TeV}$
to calculate the differential photon flux from 
Eq. \ref{diff_flux}.

However for $m_{\zp}/2 < 5 \, \rm GeV$, we have considered
the contribution of the final state radiation (FSR)
to the $\gamma$-ray signal as discussed in \cite{Cirelli:2020bpc,Essig:2013goa}.
Since we are considering one step cascade process, 
the spectrum of the emitted photon
in the rest frame of $\zp$ is given by \cite{Bystritskiy:2005ib} 
(see Appendix \ref{App:E} for the derivation)
\bea
\label{FSR_spec}
\dfrac{d N^{\rm FSR}_\gamma}{d x_\gamma}
&=&
\dfrac{\alpha_{\rm em}}{8\pi \kappa (3 - \kappa^2)}
\left[
8\left(
\dfrac{(1+\kappa^2)(3-\kappa^2)}{x_\gamma}
-2 (3-\kappa^2) + 2 x_\gamma \right) 
\ln\left(\dfrac{1+\lambda (x_\gamma)}{1-\lambda (x_\gamma)}\right)
\right.\nn\\
&&\left.
-16\left(\dfrac{(3 -  \kappa^2)(1-\kappa^2)}{x _\gamma (1 - \lambda(x_\gamma))^2} 
+  x_\gamma\right) \lambda (x_\gamma)
\right]\,\,\,.
\eea
Here $\lambda (x_\gamma) = \sqrt{1 - \dfrac{4 m_f^2}{m_{\zp}^2(1-x_\gamma)}}$, 
$\kappa^2 = 1 - \dfrac{4 m_f^2}{m_{\zp}^2}$, $m_f$ is the mass of 
final state fermions, and $x_\gamma = 2 E_\gamma/m_{\zp}$ where $E_\gamma$
is the energy of the photon in the rest frame of $\zp$.

In our analysis we have considered the FSR contribution to 
the total differential flux for $m_{\zp}/2 < 5 \, {\rm GeV}$
since for $m_{\zp}/2 \ge 5\,{\rm GeV}$ the FSR contribution is already
included in \texttt{PPPC4DMID} code.

\begin{figure}
\centering
\includegraphics[height = 6cm, width = 8cm]{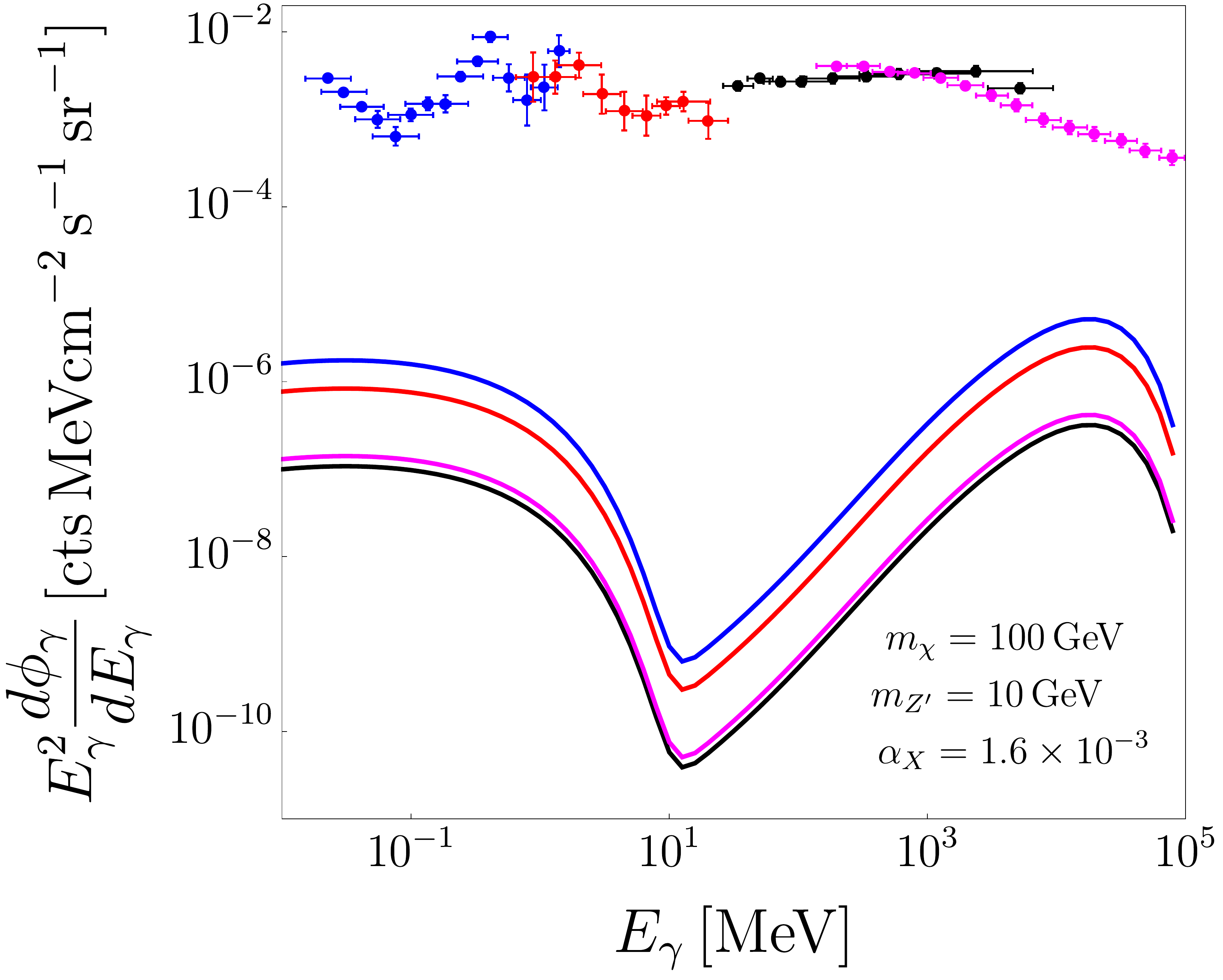}
\includegraphics[height = 6cm, width = 8cm]{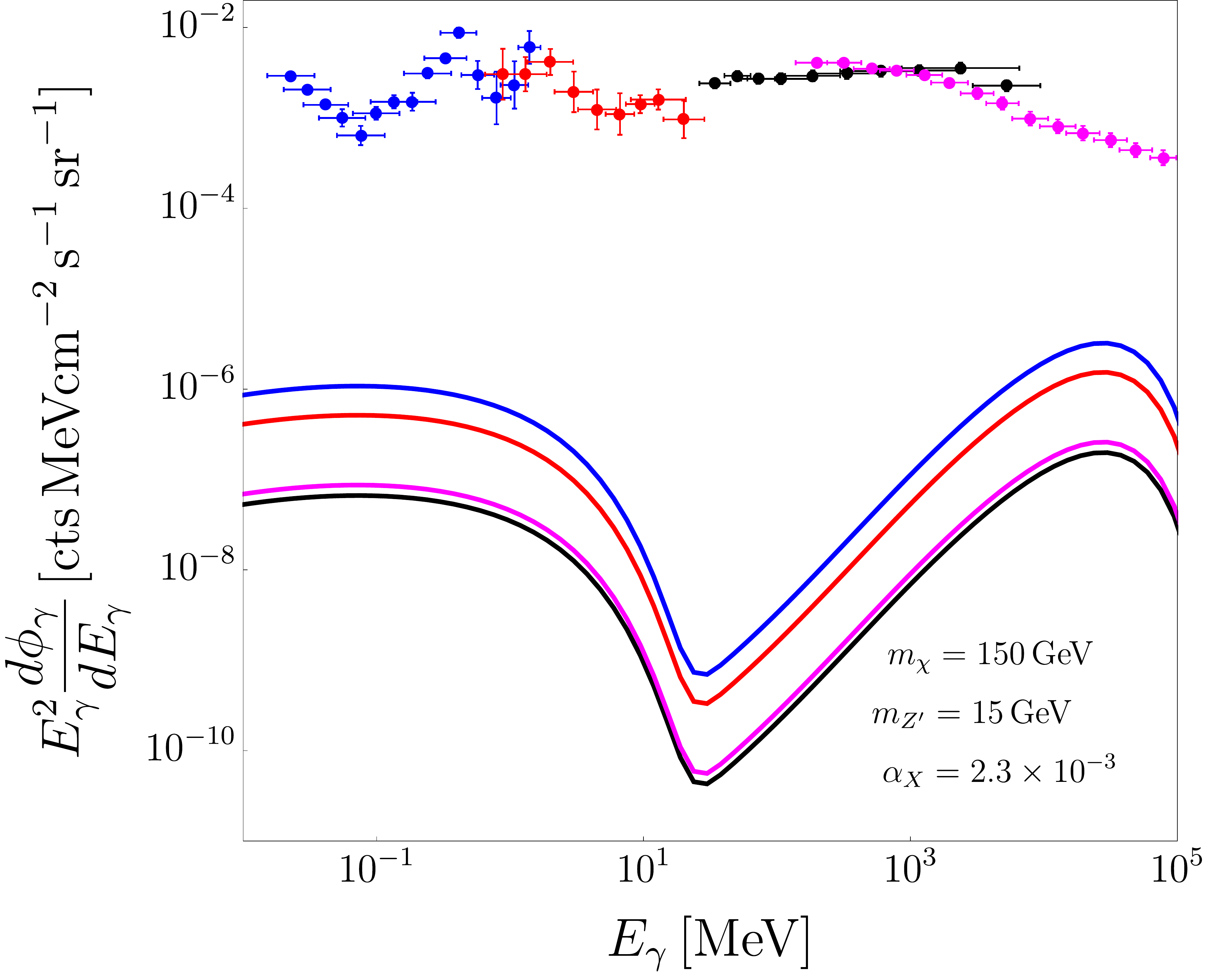}
\caption{Variation of differential photon flux as a
function of the energy of the emitted photon for two different points which
are allowed from the relic density constraint.
\textbf{\textit{Left panel:}} $E^2_\gamma$ weighted differential photon 
flux for $m_\x = 100 \, {\rm GeV}$,
$\alpha_X = 1.6 \times 10^{-3}$.
Here blue, red, black, and magenta lines represent the differential photon
flux calculated for the region of interest of INTEGRAL, COMPTEL, EGRET, and Fermi-LAT
experimental collaborations respectively whereas the experimental data
are shown by the data points with same
color code. \textbf{\textit{Right panel:}} $E^2_\gamma$ weighted differential
photon flux for $m_\x = 150 \, {\rm GeV}$ and $\alpha_X = 2.3 \times 10^{-3}$.
The color codes are same as left panel. In both the plots we have considered
$r = 10^{-1}$.
}
\label{flux}
\end{figure}

\item \textbf{Inverse Compton scattering:}
The primary high energy electrons/positrons produced from the DM annihilation
scatter with the low energy photons (such as CMB photons) and produce
high energy gamma ray which is known as Inverse Compton Scattering (ICS)
\cite{RevModPhys.42.237}. 
Following \cite{Cirelli:2009vg}, we have calculated the $\gamma$ ray flux
originating from the ICS with the ubiquitous CMB photons.
In deriving the limits we have taken the primary spectra
of $e^-, \,e^+$ from \texttt{PPPC4DMID}
for $5\,{\rm GeV}\le m_{\zp}/2 \le 100 {\rm TeV}$ 
whereas for $m_{\zp}/2< 5\, {\rm GeV}$ we have only considered the
monochromatic spectra of $e^\pm$ originating from the loop induced
$\zp \to e^+e^-$ decay.
\end{itemize}

Since $\gamma$ ray spectra from all of the above mentioned 
processes are calculated in the rest frame of $\zp$ therefore
using Eq. \ref{E9} we boost the $\gamma$ ray spectrum in
the CoM frame of DM annihilation and in this frame
the total differential $\gamma$ ray flux is given by
\bea
\dfrac{d \Phi_\gamma}{d E_\gamma}
=
\dfrac{d \Phi^{\rm prompt}_\gamma}{d E_\gamma}+
\dfrac{d \Phi^{\rm ICS}_\gamma}{d E_\gamma}\,\,,
\eea
where $\dfrac{d \Phi^{\rm prompt}_\gamma}{d E_\gamma},\,
\dfrac{d \Phi^{\rm ICS}_\gamma}{d E_\gamma}$ are 
the differential photon flux from the
prompt gamma ray and ICS respectively.

In Fig.\,\ref{flux} we show the variation of $E_\gamma^2$ weighted
differential flux as a function of the energy of the emitted photon
for two relic density satisfied benchmark points. Here the fluxes are calculated
for the region of interest (RoI) of the experimental collaboration which
are given in Table \ref{table}. As one can see from the figure
the flux for the RoI of INTEGRAL is larger than that of other RoIs. This
is because of the fact that for cuspy DM profile such as NFW density
profile the J-factor will be higher as we move towards the GC. Since
the photon flux is directly proportional to $\bar{J}_{\rm ann}$,
we get higher flux for higher values of J-factor. In Fig.\,\ref{alpha-mx},
we show the constraints from the measurement of diffuse $\gamma$-ray background
by INTEGRAL (purple region), COMPTEL (blue region),
EGRET (violet region), and Fermi-LAT (cyan region) respectively.
\begin{table}
\begin{center}
\begin{tabular}{|c|c|}
\hline
Experiments & Region of Interest (RoI)\\
\hline
INTEGRAL \cite{Bouchet:2011fn} & 
$|b| < 15^\circ,\, |l|<30^\circ$\\
\hline
COMPTEL \cite{comptel-kappadath} & $|b|< 20^\circ,\, |l|< 60^\circ$\\
\hline
EGRET \cite{Strong:2004de}& 
$20^\circ<|b| < 60^\circ,\, 0^\circ < l < 360^\circ$\\
\hline
Fermi-LAT \cite {Fermi-LAT:2012edv}
& $8^\circ < |b| < 90^\circ,\, 0^\circ< l< 360^\circ$\\
\hline
\end{tabular}
\end{center}
\caption{Region of interest of different experiments.}
\label{table}
\end{table}

\section{Constraints on the $L_\mu - L_\tau$ portal}
\label{con_zp}
\subsection{BBN constraint}
\label{BBN-constraint}
The number of relativistic degrees of freedom has been
measured from the BBN observations and this is
parametrized by defining a parameter $\rm N_{\rm eff}$.
The presence of a new light particle can alter
the number of relativistic degrees of freedom at the 
time of BBN if they are in thermal contact with the SM bath
\cite {Heeck:2014zfa,Knapen:2017xzo}.
Therefore the coupling and the mass of the new light particle
can be constrained from $\Delta N_{\rm eff} \leq 1$ at the time BBN
\cite{Mangano:2011ar}. 

In our model the presence of the dark vector boson $\zp$
can modify the value of the effective degrees of freedom in two ways:

\vspace{3pt}
\noi
i) The mediator $\zp$ can be in thermal equilibrium with the visible 
sector via $\zp \leftrightarrow \bar{\nu}_{\mu} \nu_\mu$
and $\zp \leftrightarrow \bar{\nu}_{\tau} \nu_\tau$ processes which
enhances the value of $\Delta N_{\rm eff}$,

\vspace{3pt}
\noi
ii) Due to the presence of $\zp$, neutrinos can remain in thermal equilibrium
with the SM bath via $\zp$ mediated processes 
$e^+e^- \leftrightarrow \bar{\nu}_i \nu_i$ and
$e^\pm \nu_i (\bar{\nu}_i) \leftrightarrow e^\pm  \nu_i (\bar{\nu}_i)$
where $i = \mu,\,\,\tau$. Since these processes are proportional
to $\eps^4$ (at the cross section level) therefore the effect of them
in the calculation of BBN bound is negligible
compared to the $\zp \leftrightarrow \bar{\nu}_i \nu_i$.

Therefore we have compared the reaction rate of $\bar{\nu}_i \nu_i \to \zp$
($\zp \leftrightarrow \bar{\nu}_i \nu_i$ goes out-of-equilibrium 
when the inverse decay processes $\bar{\nu}_i \nu_i \to \zp$ decouple)
with the Hubble parameter $H$ at $T = 1 \, \rm MeV$.
In Fig.\,\ref{eps-mz} the blue dashed line represents the 
contour corresponding to $\Gamma/H\vert_{T = 1\,\rm MeV} = 1$.
Besides, not to jeopardize the observations at the time BBN, the lifetime
of $\zp$ ($\tau_{\zp}$) must be smaller than 1 second and the corresponding disallowed region 
is shown in Fig.\,\ref{eps-mz} with light blue color.
\subsection{SN1987A constraint}
\label{SN1987A}
The presence of extra light degrees of freedom can enhance the
rate of cooling of SN1987A which can be constrained from
the observed neutrino luminosity after $1s$ of Supernova 1987A
\cite{Dreiner:2013mua,Chang:2016ntp}.
Since the luminosity of the neutrinos is $\mathcal{L}_\nu \simeq 
3 \times 10^{52} \, {\rm erg \, s^{-1}}$
and the mass of the core is $M_{\rm core }\sim 3 \times 10^{33} \rm g$,
the emissivity is $\mathcal{E} = \mathcal{L}_\nu/M_{\rm core}
\sim 10^{19} \, {\rm erg \, g^{-1} \, s^{-1}}$. Thus emissivity 
due to the presence of a new degree of freedom must be smaller
than $10^{19} \,{\rm erg\, g^{-1}\, s^{-1}}$ and it is known as
``Raffelt criterion" \cite{Raffelt:1996wa}. 

In our model $\zp$ can be produced from $\bar{\nu}_i \nu_i \ra \zp$
($i = \mu,\, \tau$) which can contribute to the cooling of SN1987A. Thus
we have calculated the emissivity ($\mathcal{E}_{\rm model}$) 
for these processes \cite{Escudero:2019gzq} and the corresponding expression
for the emissivity is given by
\bea
\mathcal{E_{\rm model}} &\equiv& \dfrac{1}{\rho_{\rm SN}}
\dfrac{d \rho_{\zp}}{dt} \nn\\
&=&\dfrac{3 T_{\rm SN}^3 m_{\zp} \sum_{i}\Gamma_{\zp \to \bar{\nu}_i \nu_i}}
{2 \pi^2 \rho_{\rm SN}}
\int_{y}^\infty
x \sqrt{x^2 - y^2} 
\exp(-x) \exp\left(-\dfrac{ y R_{\rm SN} \Gamma_{\zp}}{\sqrt{x^2 - y^2}}\right) d x\,\,.
\eea
Here $i = \{\mu,\,\tau,\,e\}$, $y = m_{\zp}/T_{\rm SN}$, $\Gamma_{\zp}$ is the 
total decay width of $\zp$. The other parameters such as $T_{\rm SN} = 30\rm MeV$ 
is the Supernovae (SN) temperature, the density of SN ($\rho_{\rm SN}$) 
is $3\times 10^{14} \rm g/cm^3$, and the radius of SN ($R_{\rm SN}$) is $13 \rm km$.

Now we use the condition $\mathcal{E}_{\rm model} < \mathcal{E}$
to derive the bound in the $m_{\zp}-\eps$ plane.
The grey region of Fig.\,\ref{eps-mz} is the disfavoured region 
from the SN1987A constraint. With the increase in the coupling $\eps$,
the rate of energy loss will also increase and that will put an upper bound 
on the parameter $\eps$. However, if $\eps$ is sufficiently large, then
the mean free path of $\zp$ decreases. As a result, $\zp$ produced
from the inverse decay processes will be trapped inside SN and they will
not contribute to the energy loss mechanism. Thus there will be a lower bound
on $\eps$.
\subsection{Beam dump experiments}
\label{BD}
The parameter space for a dark gauge boson
of mass $\lesssim$ $1$ GeV and kinetic mixing  $10^{-2}<\eps<10^{-8}$
can be constrained from the beam dump experiments
as discussed in \cite{Bauer:2018onh}. In our model the dark vector
boson $\zp$ can be produced from the bremsstrahlung processes
via loop induced kinetic mixing\footnote{We have not considered
the effect of $Z-\zp$ kinetic mixing and the relevant discussion
is given in footnote 2.} between $\gamma$ and $\zp$.
Following \cite{Bauer:2018onh,Bjorken:2009mm,Mo:1968cg}, 
we have derived the disallowed parameter space
in $m_{\zp}-\eps$ plane and the violet region of Fig.\,\ref{eps-mz} represents the
disallowed region.
\subsection{White dwarf cooling}
\label{WD}
The White Dwarf (WD) cooling due to emission of neutrinos can be well
described by the weak interactions. Thus any new interaction present 
in theory which contributes to the rate of cooling of WD can be
constrained from the observations \cite{Dreiner:2013tja}.

To derive the cooling constraint
due to the presence of a massive vector boson $\zp$, 
we calculate the effective Lagrangian for the new physics contribution
to the neutrino-electron interaction as discussed in \cite {Dreiner:2013tja} and 
the effective Lagrangian is given by
\bea
\label{WD1}
\mathcal{L}_{\rm NP} =- \mathcal{C}^{\zp}_{\nu_l e}
\left(\bar{\nu}_l \gamma_\mu P_L \nu_l\right)(\bar{e}\gamma^\mu e)\,\,.
\eea
Here 
\bea
\label{WD2}
\mathcal{C}^{\zp}_{\nu_l e} = 
\dfrac{\eps \sqrt{4 \pi \alpha_{\rm em}}}{m_{\zp}^2}
\Pi(0)\,\,\,
\eea
where $\Pi(0)$ is given in Eq. \ref{mixing-loop-approx}.

As discussed in \cite{Dreiner:2013tja}, 
The rate of WD cooling due to the new interaction must be smaller than the 
SM contribution to the WD cooling rate and it requires
\bea
\label{WD3}
\eps < 5 \times 10^{-5} \dfrac{m_{\zp}}{1\,{\rm MeV}}\,\,. 
\eea

Using Eq. \ref{WD3} we have derived the WD cooling constraint and it is represented
by the black dashed line in Fig.\,\ref{eps-mz}.
\subsection{Stellar cooling}
\label{stellar-cooling}
The dark vector $\zp$ can be produced inside the stellar core and
it may alter the observed rate of cooling. Therefore the properties
of $\zp$ can be constrained by requiring that the luminosity
of $\zp$ must be smaller than that of the photon luminosity.
Using $\zp$ interaction with the electromagnetic current via loop
induced kinetic mixing, we have calculated the stellar cooling
constraint as discussed in \cite{An:2013yfc,Redondo:2008aa,Hardy:2016kme}. 
The relevant parameters of the stars for calculating 
the bounds such as temperature (T), radius (R), density ($\rho$), core composition,
electron density ($n_e$), and luminosity ratio ($\mathcal{L}/\mathcal{L}_\odot$) 
are given in Table \ref{tab2} \cite{Hardy:2016kme,Dev:2020jkh}. 
In Fig.\,\ref{eps-mz} orange, red, and green regions are 
excluded from the cooling constraint of
Sun, Horizontal Branch (HB) star, and Red giant respectively.
\begin{table}
\begin{center}
\begin{tabular}{|c|c|c|c|c|c|c|}
\hline
Star & T [keV] & R [cm] & $\rho$ [$\rm g\, cm^{-3}$] & Composition 
& $n_e$ [$\rm cm^{-3}$]& $\mathcal{L}/\mathcal{L}_\odot$\\
\hline
Sun & 1& $7 \times 10^{10}$ & 150 &\makecell{25\%He\\75\%H}& $10^{26}$& 0.01\\

\hline
\makecell{Horizontal \\Branch (HB)\\stars}& 8.6 & $3.6 \times 10^{9}$& $10^4$
&${}^4 {\rm He}_2$& $3 \times 10^{27}$&5\\
\hline
Red Giant & 10 &$6 \times 10^{8}$&$10^6$ & ${}^4 {\rm He}_2$& $3 \times 10^{29}$&2.8\\
\hline
\end{tabular}
\caption{Required parameters for the calculation of stellar 
cooling limits from Sun, HB stars, and Red giant. 
Here $\mathcal{L_\odot} = 4 \times 10^{33}\, 
{\rm erg \, s^{-1}}$ is the solar luminosity.}
\label{tab2}
\end{center}
\end{table}
\subsection{Muon $g-2$ anomaly}
From the recent measurement of muon $g-2$
by Fermilab \cite{Muong-2:2021ojo}, it was found that there is a
positive deviation of $(g-2)_\mu$ from SM
prediction \cite{Davier:2017zfy,Davier:2019can,Aoyama:2020ynm}. Combining the recent result of
Fermilab with the older result of BNL E821
experiment \cite{Muong-2:2006rrc}, the experimental value of
$a_\mu \equiv \dfrac{1}{2}(g-2)_\mu$ differs from the SM prediction by
$4.2 \sigma$ and the deviation is given by \cite{Muong-2:2021ojo}
\bea
\Delta a_\mu = (251 \pm 59) \times 10^{-11}.
\eea
In our model, the dark vector boson $\zp$ can contribute to
the $(g-2)_\mu$ and the corresponding one loop integral is
given below \cite {PhysRevD.64.055006,Ma:2001md,
Banerjee:2020zvi, Fayet:2007ua,Pospelov:2008zw}.
\bea
\label{g-2}
\Delta a_\mu
&=& \dfrac{\eps^2}{4 \pi^2}
\int_0^1 \dfrac{m_\mu^2 z (1-z)^2}{m_\mu^2 (1-z)^2 + m_{\zp}^2 z} dz\,\,,
\eea
where $m_\mu$ is the mass of the muon.

\begin{figure}
\centering
\includegraphics[height = 9cm, width = 11cm]{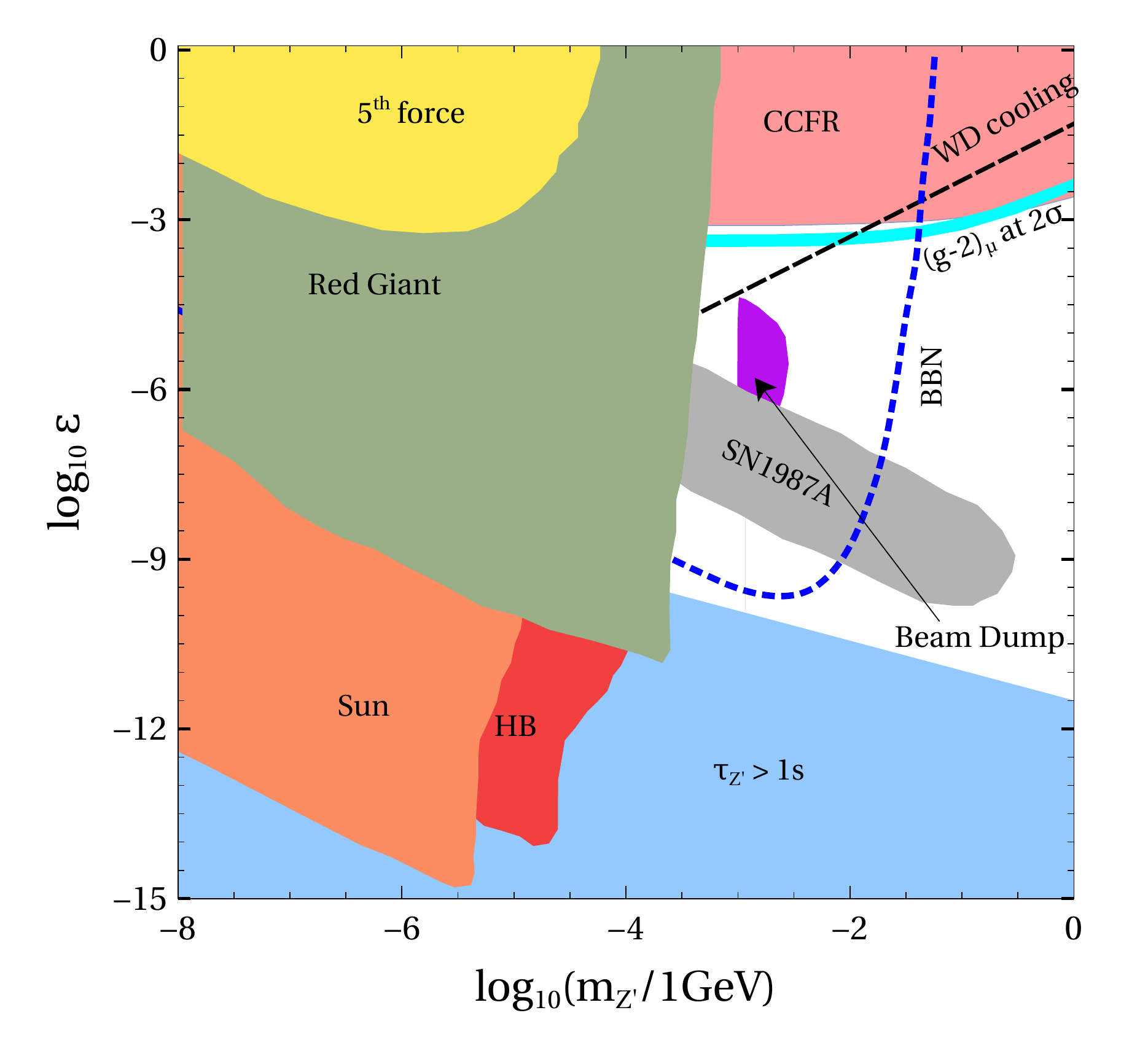}
\caption{Parameter space in $m_{\zp}-\eps$ plane for the dark gauge boson $\zp$.
The blue and black  dashed lines are the BBN and white dwarf
cooling constraint respectively. The disallowed region from $\tau_{\zp} > 1\rm s$
is shown by the light blue region. In the cyan region, $\zp$ can resolve
the Muon $g-2$ anomaly within $\pm 2\sigma$.
The grey region is disallowed from SN1987A observation,
Green, orange, and red region are disallowed from the cooling constraint of
Red Giant, Sun, and HB star. Constraint on $\zp$ from beam dump experiment
is shown by the violet region and the yellow region is
disallowed from the $5^{th}$ force experiment. The pink shaded region
is excluded (at $95\%$ C.L.) \cite{Altmannshofer:2014pba} from the 
neutrino trident production measurement at the CCFR experiment \cite{CCFR:1991lpl}.}
\label{eps-mz}
\end{figure}

Using Eq. \ref{g-2}, we have calculated the allowed region (within $\pm 2 \sigma$) of 
$m_{\zp} -\eps$ plane in which the $(g-2)_\mu$ anomaly
can be resolved by the dark vector boson $\zp$ and the corresponding
region is represented by cyan color in Fig.\,\ref{eps-mz}.

\subsection{Fifth force constraint}
\label{5th-forche}
The presence of a light $\zp$ can modify the coloumb
potential and the modification is parametrized as \cite{Jaeckel:2010ni}
\bea
V(r) = \dfrac{\alpha_{\rm em}}{r}\left(1 + |\Pi(m_{\zp}^2)|^2
\exp(-m_{\zp}r)\right)\,\,,
\eea
where $\Pi(m^2_{\zp})$ can be calculated from Eq. \ref{mixing-loop}.

As discussed in \cite{Bartlett:1988yy}, due to the modification of the Coloumb potential,
the change in Rydberg constant measurement for two different atomic 
transition ($\Delta R_{\infty}/R_\infty$) must be smaller than 
$10^{-10}$. Using this condition we have derived
the fifth force constraint and the yellow region of Fig.\,\ref{eps-mz} represents
the disallowed region from the fifth force constraint.

\section{Summary and Conclusion}
\label{conclusion}
In this work we have considered a gauged $U(1)_X$ secluded
dark sector which contains a dark vector boson $\zp$
and a Dirac fermion $\x$ which is singlet under $SU(3)_c \otimes SU(2)_L\otimes
U(1)_Y$ but charged under $U(1)_X$ gauge symmetry. We have also assumed
that the SM sector is invariant under $U(1)_{\lmt}$ gauge symmetry
and we connect visible and dark sector through the kinetic mixing
between $U(1)_X$ and $U(1)_{\lmt}$ gauge boson. We
have considered the portal coupling $\eps$ to be small enough 
so that the two sectors are thermally decoupled
but the strength of the portal coupling is sufficient to 
exchange energy between the two sectors. Therefore the dark sector
evolves non-adiabatically and in this framework we have studied
the freeze-out of dark matter which is known as ``Leak-in dark matter" scenario.
In addition, we have explored other possible mechanisms for DM production 
such as freeze-in, and reannihilation.
Since we have considered the dark sector to be thermally decoupled from the
SM bath and it is internally thermalised therefore we have also 
shown the allowed region of the model parameter space in which
both of these assumptions are valid.

Furthermore the detection prospect of our model has also been studied. 
We have investigated the constraints arising from direct detection, measurement
of diffuse $\gamma$-ray background flux by INTEGRAL, COMPTEL, EGRET, Fermi-LAT,
measurement of CMB anisotropy, and SIDM. Since there exists
a $\zp$ in our model therefore the mass and its coupling with the SM particles
i.e. $\eps$ is constrained from various laboratory and astrophysical
observations. In the light of theses observations we have also 
studied the constraints in $m_{\zp} - \eps$ plane
from BBN observations, SN1987A observations, beam dump experiment, white dwarf
and stellar cooling as well as fifth force searches. We have found that the parameter
space for the correct relic density is independent of $m_{\zp}$ 
for $m_\x/ m_{\zp} \gtrsim 10$,
which is our parameter space of interest. The allowed region from
the relic density constraint is consistent with the bounds in $m_{\zp} - \eps$
plane for $m_{\zp} > 100\,{\rm  MeV}$. Nevertheless, for $m_{\zp} \lesssim 100 \, {\rm MeV}$ there
are significant constraints coming from beam dump experiment, 
SN1987A, BBN and star cooling. The bounds from direct detection, CMB anisotropy,
SIDM, and the measurement of  diffuse $\gamma$ ray background flux
are dependent on $m_{\zp}$ as well as $m_{\x}$. We have discussed
the dependence of these bounds on the mass ratio $r = m_{\zp}/m_\x$.
We have found that for $r = 10^{-1}$
the constraints from CMB and diffuse $\gamma$-ray
observations are consistent with the relic density satisfied region
and these constraints will be relaxed for smaller values of $r$.
The SIDM and direct detection constraints for $r = 10^{-3}$
are also consistent with the allowed region from the relic
density constraint but these bounds will be significant
for smaller values of $r$.

\section{Acknowledgements}
SG would like to thank Anirban Biswas for many useful discussions during
the course of this work. SG would also like to thank University Grants Commission (UGC)
for providing financial support in the form of a senior research fellowship.
AT wishes to acknowledge the financial support provided by the
Indian Association for the Cultivation of Science (IACS), Kolkata.

\appendix
\section{$U(1)_{\lmt}$ vector portal model}
\label{App:A}
The $U(1)_{\lmt}$ vector portal model can be described
by the Eq. \ref{Lagrangian}. To express Eq. \ref{Lagrangian}
in the canonical form first we perform the following rotation.
\bea
\label{non-ortho-trans}
\begin{pmatrix}
\hat{Z}^\prime_\rho \\
\hat{Z}_{\mt_\rho}
\end{pmatrix}
 =
\begin{pmatrix}
\sec \delta & 0\\
\tan \delta & 1
\end{pmatrix}
\begin{pmatrix}
\bar{\zp}_\rho\\
\bar{Z}_{\mt_\rho}
\end{pmatrix}\,\,.
\eea
In $\left(\bar{Z}_{\mt}~~~\bar{Z}^\prime\right)^T$ basis
the gauge boson mass matrix has the following form.
\bea
\mathcal{M}^2_{\rm GB}
&=&
\hat{m}_{\mt}^2
\begin{pmatrix}
1 && \tan \delta\\
\tan \delta && \dfrac{\kappa^2}{\cos^2 \delta} + \tan^2 \delta
\end{pmatrix}\,\,\,.
\eea
Now we perform an orthogonal rotation in 
$\left(\bar{Z}_{\mt}~~~\bar{Z}^\prime\right)^T$
plane to diagonalise the gauge boson mass matrix. The orthogonal transformation
is given by
\bea
\begin{pmatrix}
\bar{Z}_{\mt_\rho}\\
\bar{Z}^\prime_\rho
\end{pmatrix}
&=&
\begin{pmatrix}
\cos \beta & -\sin \beta\\
\sin \beta & \cos \beta
\end{pmatrix}
\begin{pmatrix}
Z_{\mt \rho}\\
\zp_\rho
\end{pmatrix}\,\,\,,
\eea
where the mixing angle is
\bea
\tan 2 \beta
&=& 
\dfrac{2 \tan \delta}
{1 - \dfrac{\kappa^2}{\cos^2 \delta} - \tan^2 \delta}\,\,\,.
\eea
Therefore the diagonalised mass matrix in $\left(\zmt~~~\zp \right)^T$ basis
has the following form.
\bea
{\mathcal{M}^2_{\rm GB}}^{\rm dia.}
&=&
\begin{pmatrix}
m_{\mt}^2 & 0 \\
0 & m^{\prime 2}
\end{pmatrix}\,\,,
\eea
and the masses of the physical states are (assuming $\cos^2 \delta \simeq 1$)
\bea
m_{\mt}^2  &=& \hat{m}_{\mt}^2 \left(1 + \tan \beta \tan \delta\right)\,\,,\nn\\
m^{\prime 2} &=& \dfrac{\hat{m}^{\prime 2}}{1 + \tan \beta \tan \delta}\,\,\,.
\eea
Now we can express the mixing angle $\beta$ in terms of the physical masses
of the gauge bosons and the mixing parameter $\delta$ as
\bea
\tan \beta
&=&
\dfrac{\tan \delta}{1 - \hat{r}^2 (1 + \tan \beta \tan\delta)}\,\,,
\eea
where $\hat{r} = m^{\prime 2}/ m_{\mt}^2$.

Solving the above equation for $\tan \beta$ and choosing the condition
$\beta \to 0$ as $ \delta \to 0$ we have
\bea
\tan \beta
&=&
\dfrac{(1 - \hat{r}^2) -\sqrt{(1-\hat{r}^2)^2  - 4 \hat{r}^2 \tan^2 \delta} }{2 \hat{r}^2 
\tan \delta}\,\,.
\eea
Assuming $\delta<<1$, we have arrived at the following relations.
\bea
\tan \beta
&=&
\dfrac{\tan \delta}{1- \hat{r}^2}\,\,,\nn\\
\cos \beta
&=&
\dfrac{1 - \hat{r}^2}{\sqrt{(1-\hat{r}^2)^2 + \tan^2 \delta}}\,\,,\nn\\
\sin \beta
&=&
\dfrac{\tan\delta}{\sqrt{(1-\hat{r}^2)^2 + \tan^2 \delta}}\,\,.
\eea
Therefore we can write $\left(\hat{Z}_{\mt} ~~~ \hat{Z}^\prime\right)^T$
in terms of $\left(\zmt ~~~ \zp\right)^T$ as follows.
\bea
\hat{Z}_{\mt_\rho} &=& \left(\cos \beta + \tan \delta \sin \beta \right) Z_{\mt_\rho} - 
\left(\sin \beta - \tan \delta \cos \beta\right) \zp_\rho\,\,,\nn\\
\hat{Z}^\prime_\rho &=& \dfrac{1}{\cos \delta}
\left( \sin \beta Z_{\mt_\rho} + \cos \beta \zp_\rho\right)\,\,.
\eea

In the limit $\hat{r},\delta<<1$, we have
\bea
\hat{Z}_{\mt_\rho} &\simeq& Z_{\mt_\rho} -\hat{\eps} \zp_\rho\,\,,\nn\\
\hat{Z}^\prime_\rho &\simeq& \zp_\rho + \tan \delta Z_{\mt_\rho} \,\,\,,
\eea
where $\hat{\eps} = \hat{r}^2 \tan \delta$.
\section{Calculation of $\mathcal{C}_{{\rm SM} \to \zp} (T)$}
\label{App:B}
The collision term $\mathcal{C}_{{\rm SM \to \zp}} (T)$ 
defined in Eq. \ref{Coll_term} for a process 
like ${\rm SM}(P_1) + {\rm SM} (P_2) \to {\rm SM} (P_3) + \zp (P_4)$
(where SM denotes any SM fields) is given by
\bea
\label{coll_term_1}
\mathcal{C}_{{\rm SM \to \zp}} (T)
&=&
\int d \Pi_i E_4
(2 \pi)^4 \delta^4 (P_1 + P_2 - P_3 -P_4)
\overline{|\mathcal{M}|^2}
f_{\rm SM} (p_1, T) f_{\rm SM}(p_2, T)\,\,,
\eea
where the definition of all the quantities used in this equation
are same as the definitions given in Eq. \ref{Coll_term}.
Since SM fields are in thermal equilibrium therefore
we use $f_{\rm SM}(p_i,T)$ to be Maxwell-Boltzmann (MB)
distribution. Using the MB distribution function
for SM fields and the energy conservation,
the above equation takes the following form.
\bea
\label{coll_term_2}
\mathcal{C}_{{\rm SM \to \zp}} (T)
&=&
\int d \Pi_3 d \Pi_4 E_4 \exp\left(-\dfrac{E_3 + E_4}{T}\right) 
4 E_3 E_4 v \sigma _{3 4 \to 12}\,\,,
\eea
where $\sigma_{34 \to 12}$ is the annihilation cross section
of the process $ {\rm SM} (P_3) + \zp (P_4) \to {\rm SM} (P_1) + {\rm SM} (P_2)$
and $4 E_3 E_4 v$ is the usual flux factor. Following the technique given in 
\cite{Gondolo:1990dk} we can write Eq. \ref{coll_term_2} as
\bea
\label{coll_term_3}
\mathcal{C}_{{\rm SM \to \zp}} (T)
&=&
\dfrac{g_3 g_4 \pi^2}{4 (2 \pi)^6}
\int_\Upsilon
^\infty d \,\hat{s}
\int_{\sqrt{\hat{s}}}^\infty d\, E_+
\int_{r_1}^{r_2} d\, E_-
\left[\left(E_+^2 - E_-^2\right)(E_+ - E_-)\right.\nn\\
&&
\left.~~~~~~~~~~~~~~~~~~~~~~~~~~~~~~~~~~~~~~~
\times
(\sigma v)_{34\to12}\exp\left(-\dfrac{E_+}{T}\right)
\right]\,\,,
\eea
where we define $E_\pm = \left(E_3 \pm E_4\right)/2$, $r_1$ and $r_2$
are the lower and upper limit of $E_-$ respectively and 
the lower limit $\Upsilon$ of $\hat{s}$ integration are defined as follows.
\bea
\Upsilon &=& {{\rm Max}\left[(m_1 + m_2)^2, (m_3 + m_{\zp})^2\right]}\,\,,\nn\\
r_1 &=& E_+ \,\eta - 2 \sqrt{E_+^2 - \hat{s}}\,\lambda\,\,\,,\nn\\
r_2 &=& E_+ \,\eta + 2 \sqrt{E_+^2 - \hat{s}}\,\lambda\,\,\,,\nn\\
\text{where}\nn\\
\lambda &=& \dfrac{\sqrt{\hat{s} - (m_3 + m_4)^2}
\sqrt{\hat{s}- (m_3^2 - m_4)^2}}{2 \hat{s}}\,\,,\nn\\
\eta&=& \dfrac{m_3^2 - m_4^2}{\hat{s}}\,\,.
\eea
Thus after performing the integration over $E_+$ and $E_-$
in Eq. \ref{coll_term_3} we can write the final 
form of the collision term as
\bea
\label{coll_term_final}
\mathcal{C}_{{\rm SM}\to \zp} (T)
&=&
\dfrac{g_3 g_4 T}{ (2 \pi)^4}
\int_\Upsilon^\infty
\lambda^2\,\hat{s}^2\,\left(1 - \eta \right)
\sigma_{34\to 12} \, K_2 \left(\dfrac{\sqrt{\hat{s}}}{T}\right)
d \,\hat{s}\,\,\,\,.
\eea
\section{Calculation of the reaction rate ($\Gamma$)}
\label{App:C}
As discussed in section \ref{wimp-next-door}, to identify the relevant 
parameter space for the thermally decoupled dark sector
we need to compare the total reaction rate for all
processes discussed  in section \ref{DS_temp} with the hubble parameter.

For a process $ A(P_1) +\zp (P_2) \to B (P_3) + C(P_4)$ where
$A, B, C$ are the SM particles, the reaction
rate per $\zp$ particle is given by
\bea
\Gamma_{A \zp \to B C}
= n^{\rm eq}_A (T) \langle \sigma v \rangle_{A \zp \to B C}\,\,
\eea
where $n^{\rm eq}_A (T)$ is the equilibrium number density of A and 
the thermally averaged cross section of the above mentioned
process can be written as follows \cite{Gondolo:1990dk}.
\bea
\langle \sigma v \rangle_{a \zp \to b c}
= \dfrac{g_A {g_\zp}}{n_A^{\rm eq} (T) n^{\rm eq}_{\zp}(T)}
\dfrac{T}{2 (2 \pi)^4}
\int_\Upsilon^\infty
\sigma(\hat{s})_{A \zp \to B C} \, \tilde{\Gamma}(\hat{s},m_A^2,m_{\zp}^2) 
\, {\rm K}_1 \left(\dfrac{\sqrt{\hat{s}}}{T}\right) d \hat{s}\,\,,
\eea
where 
\bea
\Upsilon &=& {\rm Max}\left[(m_A + m_{\zp})^2, (m_B + m_C)^2\right]\,\,\nn\\
\tilde{\Gamma}&=& \dfrac{\hat{s}^2 - 2 \hat{s} (m_A^2 + m_{\zp}^2) + 
(m_A^2 - m_{\zp}^2)^2}{\sqrt{\hat{s}}}\,\,.
\eea
Therefore the total reaction rate per $\zp$ is defined as
\bea
\Gamma = \sum_{\text{All channels}} \Gamma_{A \zp \to B C}\,\,,
\eea
where the summation is taken over all the channels discussed in section \ref{DS_temp}.
\section{Thermalisation of the dark sector}
\label{App:D}
In our analysis for the evolution of the dark sector, we have
assumed that the dark sector is internally thermalised i.e.
the DM $\x$ and dark vector boson $\zp$ is in thermal equilibrium.
Therefore it is important to validate that the initial
number density of $\x$ and $\zp$ produced from the SM bath and
their interaction strength are sufficient to keep them in
thermal equilibrium with different temperature from SM.

To study the allowed parameter space for the internal thermalisation
first we study the production of $\x$ and $\zp$ from SM bath. The Boltzmann
equation for the production of $\x$ and $\zp$ are as follows.
\bea
\label{D1}
\frac{d n_{\x}}{dt} + 3 H n_{\x}
&=& \sum_f \mathcal{C}_{\bar{f}f \to \bar{\x}\x} (T)\,\,\,,\nn\\
\frac{d n_{\zp}}{dt} + 3 H n_{\zp}
&=& \mathcal{C}_{{\rm SM} \to \zp} (T)\,\,\,.
\eea
Here $H$ is the Hubble parameter defined in Eq. \ref{Hubble}, 
$\mathcal {C}_{\bar{f}f \to \bar{\x}\x (T)}$ is the collision
term for the DM production from SM bath and $f = \mu, \tau, \nu_\mu, \nu_\tau$.
The sum of the collision terms for the $\zp$ production 
from the processes mentioned in section \ref{DS_temp} is denoted by
$\mathcal{C}_{{\rm SM} \to \zp } (T)$. 

Now we define the co-moving number density of $\x$ and $\zp$ as 
$Y_\x = n_\x/s$ and $Y_{\zp} = n_{\zp}/s$ where $s$ is the 
entropy density of the Universe. Now using the definition
of $Y_\x$ and $Y_{\zp}$ we can write Eq. \ref{D1} as follows.

\bea
Y_{\x} (T) = \sum_f \int_{T}^{T_0} 
\dfrac{\mathcal{C}_{\bar{f}f \to \bar{\x}\x} (\bar{T})}
{\bar{T} H(\bar{T})s (\bar{T})}d \bar{T}\,\,\,,\nn\\
Y_{\zp} (T) = \int_{T}^{T_0}
\dfrac{\mathcal{C}_{{\rm SM}\to \zp} (\bar{T})}
{\bar{T} H(\bar{T})s (\bar{T})}d \bar{T}\,\,\,,
\eea
where $T_0$ is the temperature of the Early Universe and $T_0 >> T$.

\begin{figure}
\centering
\includegraphics[height = 8cm, width = 8cm]{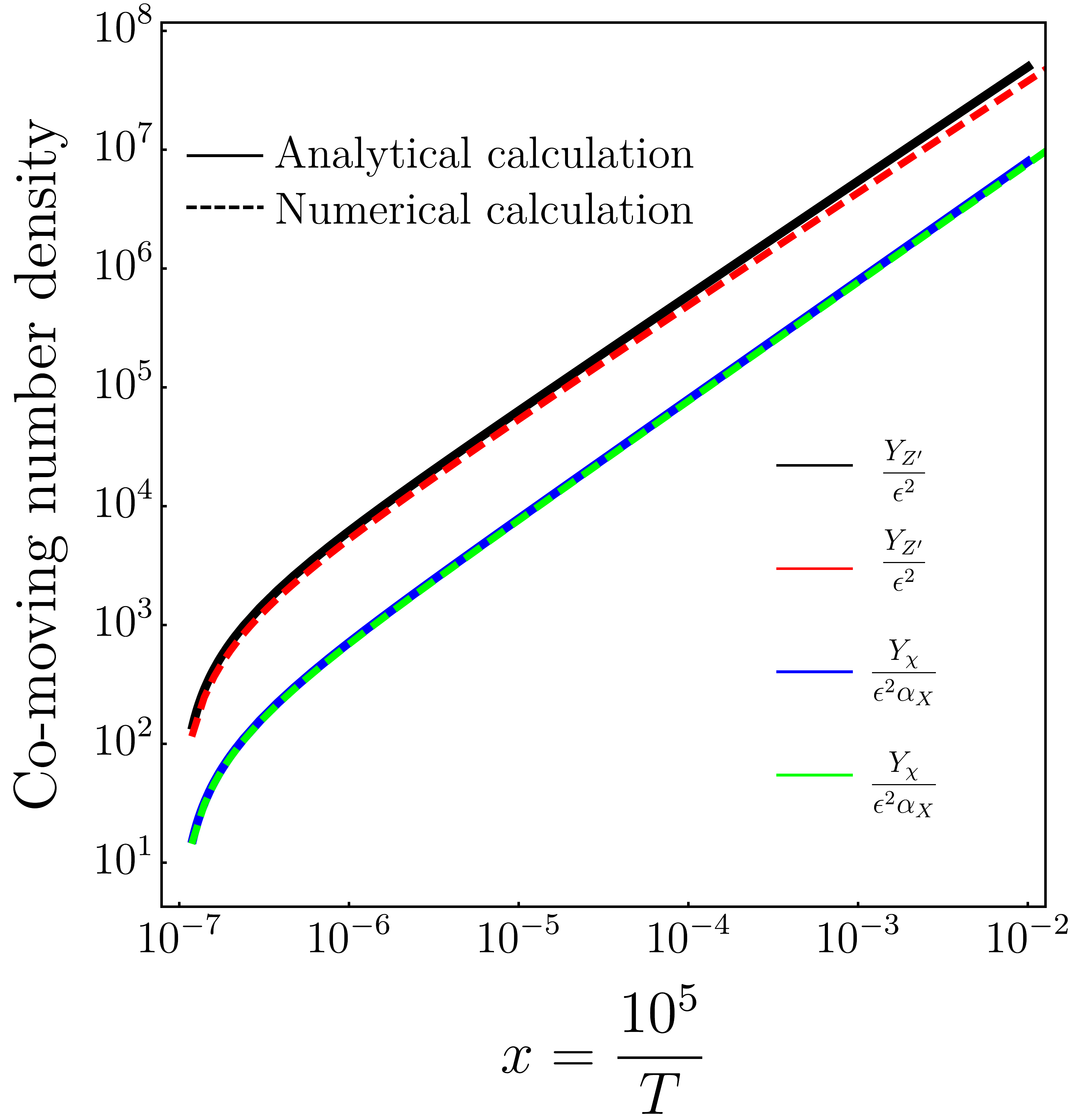}
\caption{Numerical and analytical calculation of $Y_{\x}$ and $Y_{\zp}$.}
\label{int_therm_1}
\end{figure}

In the early Universe we have assumed $m^2/\hat{s} <<1$ where $\sqrt{\hat{s}}$
is the total energy of the initial state particles and $m$ represents the
masses of the particles in the scattering process. Under this assumption
we have calculated the collision terms analytically and in 
Fig.\ref{int_therm_1} we have compared the result from the analytical 
and full numerical calculations. As one can see 
from the figure, the analytical estimate
of $Y_\x$ and $Y_{\zp}$ are consistent with the full numerical calculation
therefore from now on we will use the analytical 
results of $Y_\x$ and $Y_{\zp}$ for the remaining part of this section.

\begin{figure}
\centering
\includegraphics[height = 7cm, width = 8cm]{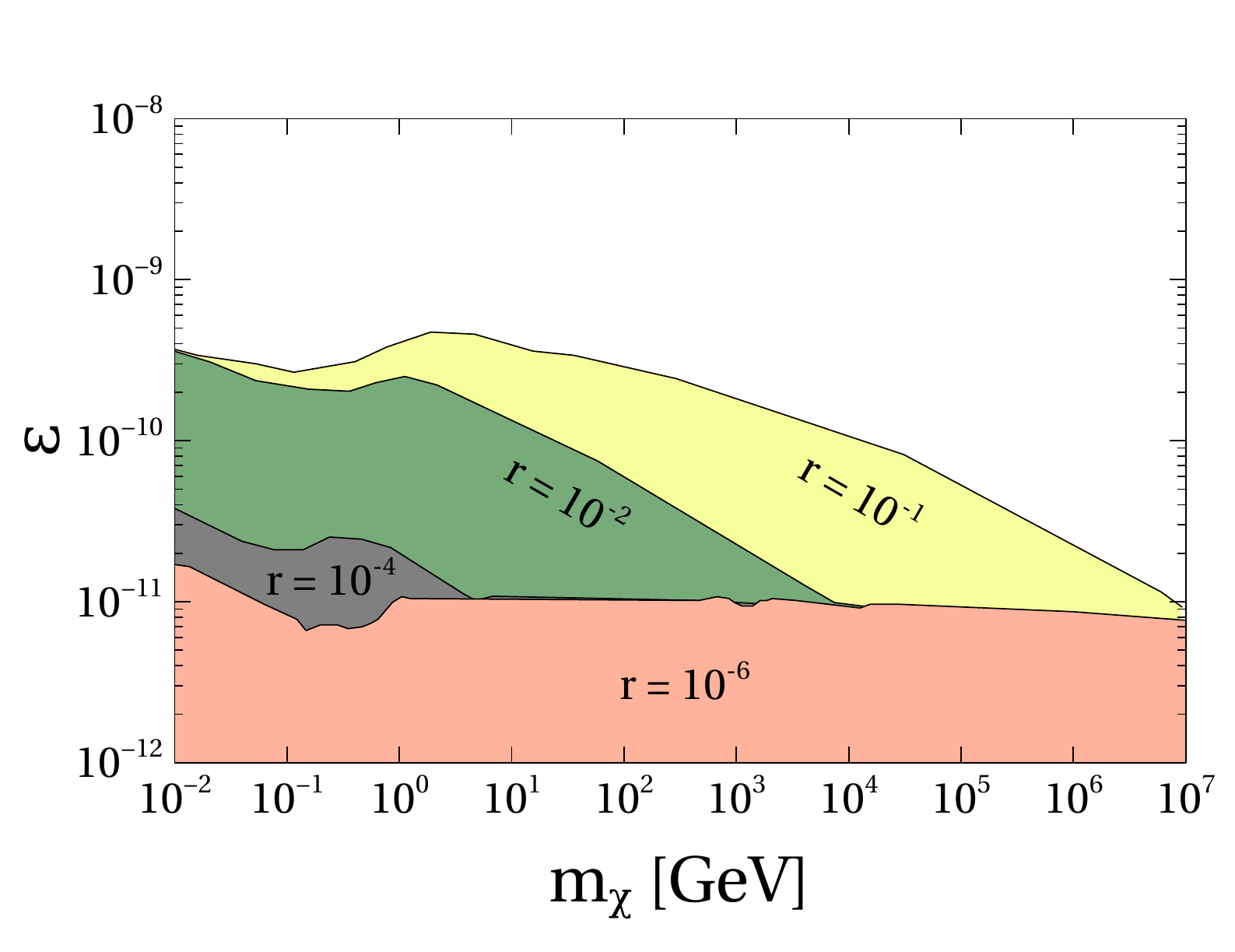}
\caption{Disallowed region from the thermalisation criteria in the $m_\x - \eps$
plane for different values of $r$. The disallowed regions for
$r = 10^{-1},\,10^{-2},\,10^{-4}$, and $10^{-6}$
are denoted by yellow, green, grey, and
pink color respectively. }
\label{Int_therm}
\end{figure}

Thus to check the thermalisation of the dark sector 
by $\bar{\x} \x \to \zp \zp$ process we need to 
calculate the region in which $\Gamma_{2\to2}<H$, where
we define $\Gamma_{2\to2}$ as
\bea
\Gamma_{2\to2} = s(T)\,{\rm Max} 
\left[ Y_\x (T )\langle\sigma v \rangle_{\bar{\x} \x \to Z^\prime Z^\prime},\,
Y_{\zp} (T) \langle\sigma v \rangle_{Z^\prime Z^\prime \to \bar{\x} \x}
\right]\,\,.
\eea

However $2\to3$ processes such as $\x X \to \x X \zp$ $(X = \bar{\x}, \zp)$
can also thermalise the dark sector as discussed in \cite{Garny:2018grs}. Though
these $2\to3$ processes are suppressed by another extra vertex factor 
$\alpha_X$ in comparison to the $2\to2$ processes but for soft momentum exchange
(i.e. the exchanged momentum in the propagator $\sim \sqrt{4 \pi \alpha_X} \tp$) the 
rate of these $2\to3$ processes are comparable to the rate of $2\to2$ process. 
Since the time scale of both the processes are comparable therefore effect
of several $2\to2$ collision before the emission of $\zp$ has to be taken
into account. The emission rate of $\zp$ is significantly modified
due to the presence of multiple $2\to2$ scattering and it is known
as Landau-Pomeranchuk-Migdal (LPM) effect \cite{Migdal:1956tc,Landau:1953gr}.

Therefore following \cite{Garny:2018grs,Arnold:2002zm}, 
we have calculated the reaction rate
of $2\to3$ processes ($\Gamma_{2\to3}$) and derive the disallowed 
region from the thermalisation criteria from the following relation.
\bea
\Gamma_{2\to3} + \Gamma_{2\to2} <H.
\eea

The disallowed region in the $m_\x-\eps$ plane in which
the dark sector is not internally thermalised, is shown in
Fig.\,\ref{Int_therm}. In this figure we have considered
four different values of $r$ which is defined as $r = m_{\zp}/m_\x$.
In deriving the limit, we use Eq. \ref{xi-approx}, Eq. \ref{relic_approx}, 
and Eq. \ref{cross-swave}
to express $\alpha_X$ in terms of $m_\x,\,m_{\zp}$,
and $\eps$.
\section{Photon spectrum for final state radiation}
\label{App:E}
\begin{figure}
\centering
\includegraphics[scale = 0.4]{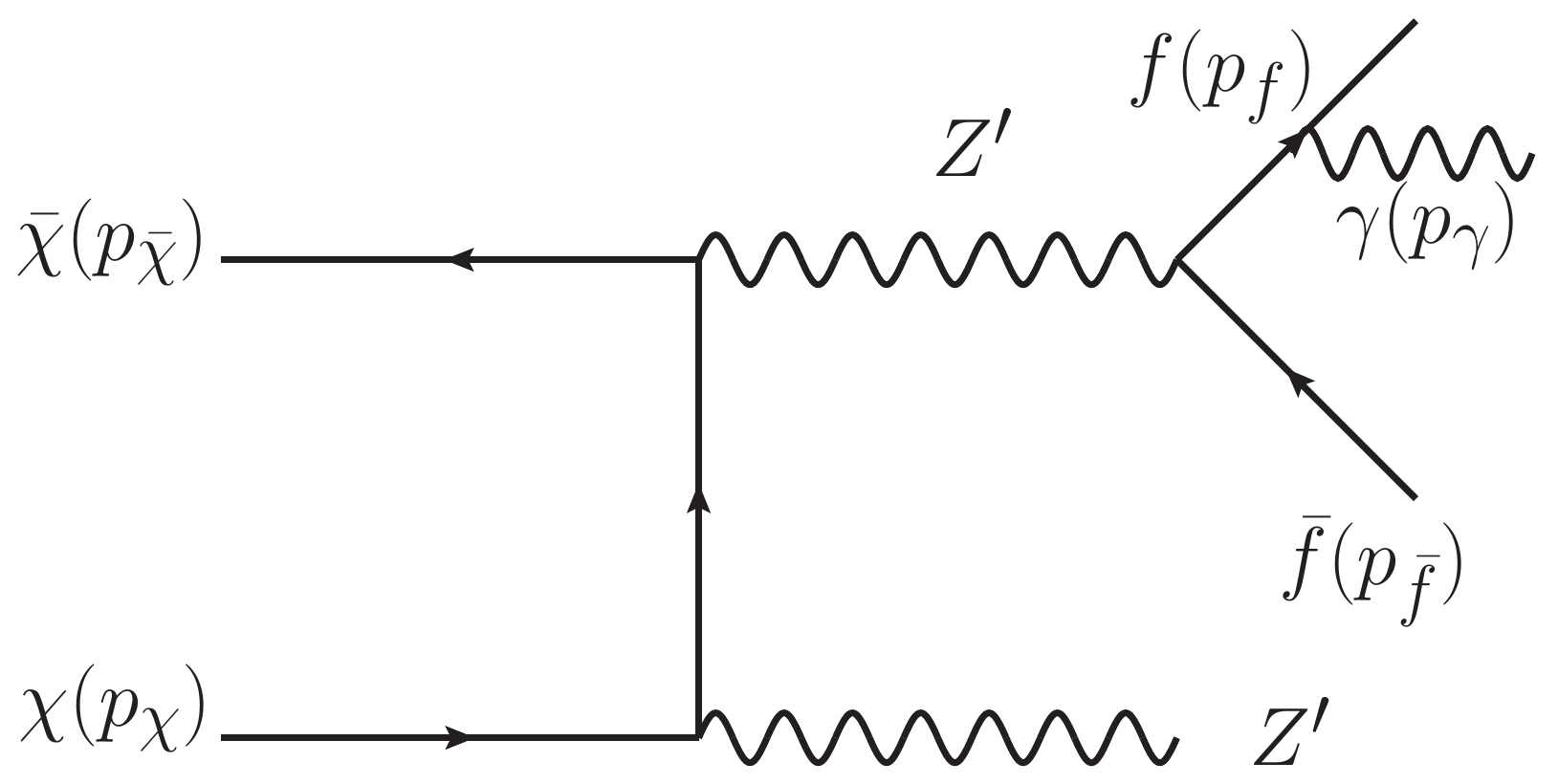}
\includegraphics[scale = 0.4]{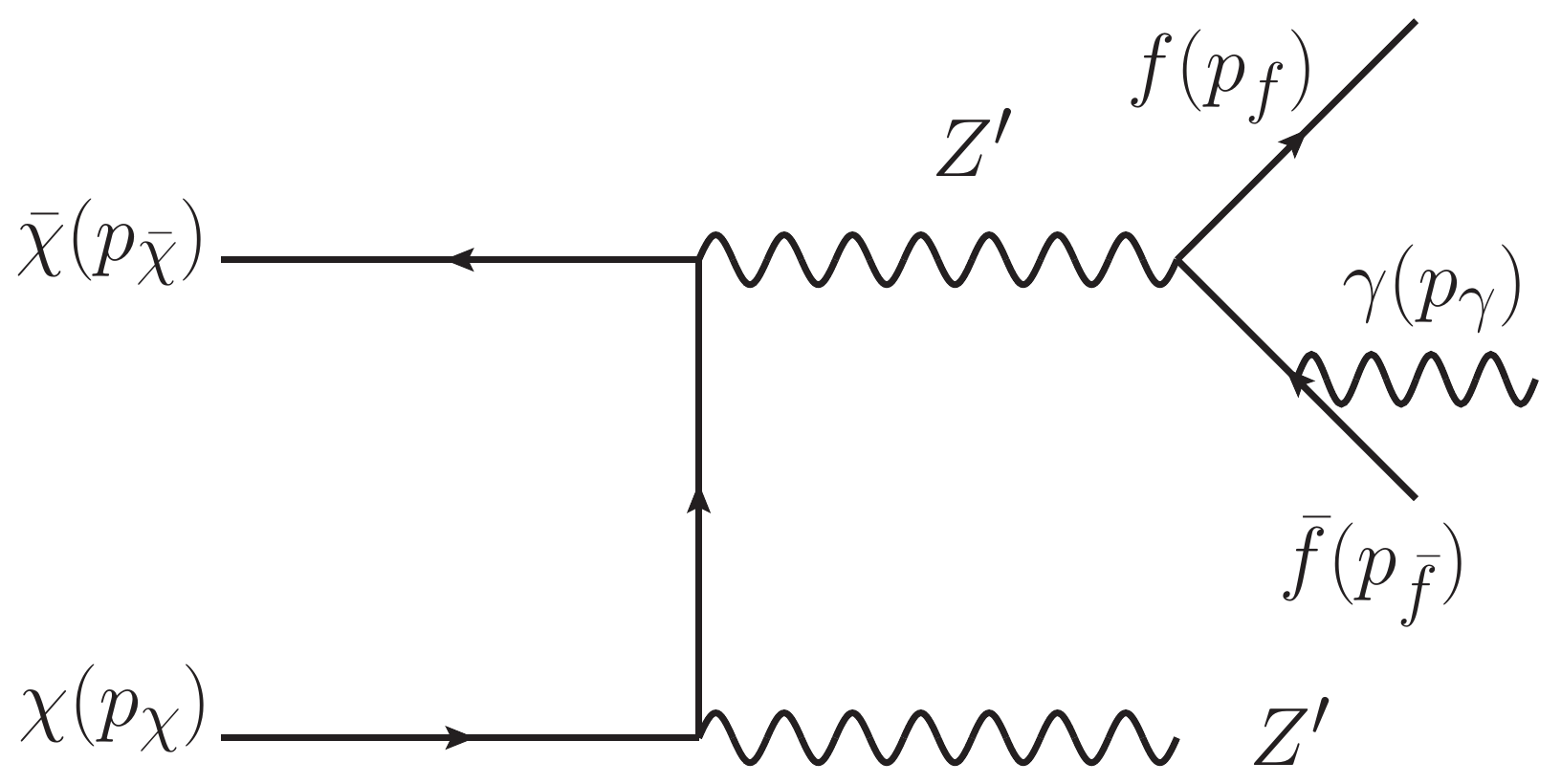}
\caption{Feynman diagrams for final state radiation}
\label{FSR-dia}
\end{figure}

In our scenario the final state radiation occurs
from one step cascade process shown in Fig.\,\ref{FSR-dia}.
The $\gamma$ ray spectrum in the rest frame of
$\zp$ is given by
\bea
\label{E1}
\dfrac{d N^{\rm FSR}_\gamma}{d x_\gamma}
&=& \dfrac{1}{\Gamma_{\zp \to \bar{f}f}}
\dfrac{\alpha_{\rm em} \eps^2 m_{\zp}}{192 \pi^2}
\int_{-1}^{1}\dfrac{d \cos \theta}{2}
\int_{x_{f_{\rm min}}}^{x_{f_{\rm max}}}
d x_f \mathcal {G}(x_\gamma, x_f, m_f, {m_\zp})\,\,,
\eea
where $x_f = 2 E_f/m_{\zp}$, $x_\gamma = 2 E_\gamma/m_{\zp}$. $E_\gamma$,
$E_f$, and 
$\theta$ are the energy of the photon, energy of the fermion $f$, and
angle of emission of $\gamma$ respectively and all of these quantities
are measured in the rest frame of $\zp$. The
upper and lower limit of the $x_f$ integration are
\bea
x_{f_{\rm min}} = \dfrac{1}{2}\left[(2-x_\gamma) - 
x_\gamma \sqrt{1-\dfrac{4 m_f^2}{m_{\zp}^2 (1 - x_\gamma)}
}\right]\,\,\,,\nn\\
x_{f_{\rm max}} = \dfrac{1}{2}\left[(2-x_\gamma) +
x_\gamma \sqrt{1-\dfrac{4 m_f^2}{m_{\zp}^2 (1 - x_\gamma)}
}\right]\,\,\,.
\eea

The decay
width of $\zp$ into a pair of SM fermions is
\bea
\label{E2}
\Gamma_{\zp \to \bar{f}f}
&=& \dfrac{\eps^2 m_{\zp}}{12 \pi}\sqrt{1 - \dfrac{4 m_f^2}{m^2_{\zp}}}
\left(1 + \dfrac{2 m_f^2}{m^2_{\zp}}\right)\,\,.
\eea

Now in Eq. \ref{E1} the function $\mathcal{G}(x_\gamma, x_f, m_f, m_{\zp})$ is  
given by
\bea
\label{E3}
\mathcal{G} (x_\gamma, x_f, m_f, m_{\zp})
&=& g_{\mu \nu} X^{\mu \nu}\,\,\,,
\eea
where
\bea
\label{E4}
X^{\mu \nu} = Tr&&\left[
\left(\gamma_\alpha \dfrac{\slashed{p}_f + \slashed{p}_\gamma 
+ m_f}{2 p_f . p_\gamma} \gamma^\mu
+
\gamma^\mu \dfrac{-\slashed{p}_{\bar{f}} - \slashed{p}_\gamma 
+ m_f}{2 p_{\bar{f}} . p_\gamma} \gamma_\alpha\right)
\left(\slashed{p}_{\bar{f}} -m_f\right)\right.\nn\\
&& \left.
\left(\gamma^\nu \dfrac{\slashed{p}_f + \slashed{p}_\gamma
+ m_f}{2 p_f . p_\gamma} \gamma^\alpha
+
\gamma^\alpha \dfrac{-\slashed{p}_{\bar{f}} - \slashed{p}_\gamma
+ m_f}{2 p_{\bar{f}} . p_\gamma} \gamma^\nu\right)
\left(\slashed{p}_f + m_f\right)
\right]\,\,\,.
\eea

Using $E_{\bar{f}} = \dfrac{m_{\zp}}{2}(2 - x_\gamma - x_f)$ 
in Eq. \ref{E4}, we can write the final form of photon 
spectrum in the rest frame of $\zp$ 
from Eq. \ref{E1} as follows.
\bea
\label{E5}
\dfrac{d N^{\rm FSR}_\gamma}{d x_\gamma}
&=& \dfrac{\alpha_{\rm em}}{8 \pi \kappa (3-\kappa^2)}
\left[\mathcal{A}_1 \ln \left(\dfrac{1 + \lambda (x_\gamma)}{1 - \lambda (x_\gamma)}
\right) - \mathcal{A}_2\right]\,\,\,,
\eea
where
\bea
\label{E6}
\kappa&=& \sqrt{1 - \dfrac{4m_f^2}{m_{\zp}^2}}\,\,,\nn\\
\lambda (x_\gamma)&=& \sqrt{1 - \dfrac{4 m_f^2}{m^2_{\zp} (1 - x_\gamma)}}\nn\,\,\,,\\
\mathcal{A}_1
&=&8\left[ \dfrac{(1 + \kappa^2)(3 - \kappa^2)}{x_\gamma}
-2 (3 - \kappa^2) + 2 x_\gamma
\right]\nn\,\,\,,\\
\mathcal{A}_2 &=&16 \lambda(x_\gamma) \left[
\dfrac{(1-\kappa^2)(3-\kappa^2)}{x_\gamma (1 - \lambda(x_\gamma)^2)} + x_\gamma
\right]\,\,.
\eea

In the limit $\dfrac{4 m_f^2}{m_{\zp}^2} \to 0$, Eq. \ref{E5} takes
the following form.
\bea
\label{FSR}
\dfrac{d N^{\rm FSR}_\gamma}{d x_\gamma}
&\simeq&
\dfrac{\alpha_{\rm em}}{\pi}
\left(
\dfrac{1 + (1 - x_\gamma)^2}{x_\gamma}
\right)
\left[
\ln \left(\dfrac{m^2_{\zp} (1 - x_\gamma)}{m_f^2}\right) - 1
\right]\,\,\,.
\eea

Finally to calculate the $\gamma$ ray spectrum in the centre of mass (CoM) frame
of DM annihilation \cite{Elor:2015tva} we can write from Eq. \ref{E1}
\bea
\label{E7}
N^{\rm FSR}_\gamma
&=&
\int_{-1}^1 \dfrac{d \cos \theta}{2} \int d x_\gamma \int d x_f
\mathcal{G}(x_\gamma, x_f, m_f, m_{\zp})\,\,.
\eea
Now we use the following property of Dirac delta function.
\bea
\label{E8}
\int d E^0_\gamma 
\delta\left(E^0_\gamma-\gamma_{\zp} E_\gamma 
(1 + \beta_{\zp}\cos \theta)\right) = 1\,\,\,,
\eea
where $E^0_\gamma$ is the photon energy in the CoM frame of DM annihilation,
$\gamma_{\zp} = \sqrt{s_0}/2 m_{\zp}$, $\beta_{\zp} = \sqrt{1 - \gamma^{-2}}$,
and $\sqrt{s_0}$ is the total energy of the annihilating DM particles
in the CoM frame.

Using Eq. \ref{E8} in Eq. \ref{E7} we can write the photon
spectrum in the CoM frame of DM annihilation as follows.
\bea
\label{E9}
\dfrac{d N^{\rm FSR}_\gamma}{d x^0_\gamma}
&=&
\dfrac{2}{\beta_{\zp}}
\int_{x_{\gamma_{\rm min}}}^{x_{\gamma_{\rm max}}}
\dfrac{1}{x_\gamma} \dfrac{d N^{\rm FSR}_\gamma}{d x_\gamma} d x_\gamma\,\,\,,
\eea
where $x^0_\gamma = 2 E^0_\gamma/\sqrt{s_0}$, 
$x_{\gamma_{\rm min}} = \dfrac{2 x^0_\gamma (1 - \beta_{\zp})}{\gamma_{\zp}^{-2}}$, and 
$x_{\gamma_{\rm max}} = {\rm Max}\left[1, 
\dfrac{2 x^0_\gamma (1 +  \beta_{\zp})}{\gamma_{\zp}^{-2}}\right]$. 
The 2 factor in Eq. \ref{E9} arises due to production of two $\zp$
from each DM annihilation process.

\bibliographystyle{JHEP}
\bibliography{References}
\end{document}